\documentclass[aps,prb,superscriptaddress,twocolumn,no-footinbib,showpacs,floatfix,amsmath,amssymb]{revtex4-1}

\usepackage{graphicx}
\usepackage{dcolumn}
\usepackage{bm}
\usepackage[normalem]{ulem}
\usepackage{subfigure} 
\usepackage{soul,xcolor}
\usepackage{siunitx}
\usepackage{textcomp}
\setstcolor{red}

\begin{document}

\title{Enhanced asymmetric valley scattering by scalar fields in non-uniform out-of-plane deformations in graphene}

\author{Ramon Carrillo-Bastos}
\email{ramoncarrillo@uabc.edu.mx}
\affiliation{Facultad  de  Ciencias,  Universidad  Aut\'onoma  de  Baja  California, 22800  Ensenada,  Baja  California,  M\'exico.}
\author{Marysol Ochoa}
\affiliation{Departamento de F\'isica T\'eorica - CNyN, Universidad Nacional Aut\'onoma de M\'exico, Apdo. Postal 14, 22800 Ensenada, Baja California, M\'exico}
\affiliation{Plantronics Inc. Blvd. Bellas Artes No.20308., Tijuana, Baja California, M\'exico, 22444}
\author{Sa\'ul A. Zavala}
\affiliation{Tecnol\'ogico Nacional de M\'exico/I.T. Ensenada. Departamento de Ciencias B\'asicas. Boulevard Tecnol\'ogico No. 150, Ex-ejido Chapultepec, Apdo. Postal 22780 Ensenada, Baja California, M\'exico}
\affiliation{Facultad  de  Ingenier\'ia y Dise\~no,  Universidad  Aut\'onoma  de  Baja  California, 22800  Ensenada,  Baja  California,  M\'exico.}
\author{Francisco Mireles}
\affiliation{Departamento de F\'isica T\'eorica - CNyN, Universidad Nacional Aut\'onoma de M\'exico, Apdo. Postal 14, 22800 Ensenada, Baja California, M\'exico}
\date{\today}

\begin{abstract}
We study the electron scattering produced by local out-of-plane strain deformations in the form of Gaussian bumps in graphene. Of special interest is to take into account the scalar field associated with the redistribution of charge due to deformations, and in the same footing as the pseudomagnetic field. Working with the Born approximation approach we show analytically that even when a relatively small scalar field is considered, a strong backscattering and enhancement of the valley splitting effect could arise as a function of the energy and angle of incidence. In addition, we find that the valley polarization can reverse its sign  as the incident energy is increased. These behaviors are totally absent if the scalar field is neglected or screened. Interestingly, we find that there is a further possibility of controlling the valley scattering polarization purely by electrical means through the presence of external scalar fields in combination with strain fields. These results are supported by quantum dynamical simulations of electron wave packets. Results for the average trajectories of wave packets in locally strained graphene clearly show focusing and beam splitting effects enhanced by the presence of the scalar field that can be of interest in the implementation of valleytronic devices.
\end{abstract}

\maketitle

\section{Introduction}
The appearance in graphene of massless Dirac Fermions and constant velocity $v_F$ at low energies,  emerges because its two equivalent carbon sublattices of trigonal symmetry\cite{neto2009electronic}.  In the presence of strain, the corresponding graphene Hamiltonian and thereby its linear dispersion laws near the $K$($K^{'}$) Dirac points gets modified accordingly. From the theoretical point of view, symmetry considerations allows up to six additional terms in its low energy Hamiltonian\cite{vozmediano2010gauge,AMORIM20161,naumisReview,RAMEZ2013}. Namely, those terms due to uniform strains which   give rise to the pseudomagnetic and scalar fields, a gap opening term due to possible non-uniform strains, a  Dirac cone strain-induced tilt term, and those due to the presence of isotropic and anisotropic position dependent Fermi velocities. Among them,  the strain-induced pseudomagnetic field effects associated with the shift of the Dirac cones in the momentum space\cite{Caio} is the one most studied recently. The latter because of its natural interpretation as a sort of magnetic field\cite{Field-Kane-Mele,vozmediano2010gauge} (pseudomagnetic field)  that under appropriate physical conditions generates, by analogy with a real magnetic field, a  Landau level spectrum\cite{Theory-LL01}, phenomena that has been beautifully demonstrated in recent experiments\cite{Crommie}. Moreover, in the same manner as the real magnetic field couples with the intrinsic angular momentum of the electron, the pseudomagnetic field can also couple with the pseudospin\cite{sasaki2008pseudospin}, generating a Zeeman-like splitting as observed in very recent STM experiments\cite{georgi2017tuning}. 

Several experimental setups from different groups have  reported to produce strains in graphene membranes\cite{AMORIM20161,naumisReview}. They range from the deposition on  substrates\cite{jiang2017,lim2015structurally,Crommie}, the formation of bubbles\cite{bunch2008,khestanova2016}, generation of deformations by STM\cite{klimov,georgi2017tuning}  or AFM\citep{nemes2017preparing} tips, to the deposition of graphene membranes in nanostructured arrays\cite{tomori,zhang2017}. The induced pseudomagnetic field in graphene has several advantages over the real magnetic field, for instance, since graphene is very flexible\cite{AMORIM20161}  the magnitudes of the pseudomagnetic field obtained by strain are many times stronger compared with real magnetic fields\cite{Crommie,georgi2017tuning} ($\sim 300$ T). 

On the other hand, due to its mechanical origin,  the pseudomagnetic field does not break time-reversal symmetry; so in the effective Hamiltonian it appears only as a reverse sign in the $K$($K^{'}$)  valleys\cite{PereiraPRL}. This unique physical characteristic have been proposed as a mechanism of control of the valley degree of freedom in various scenarios. For instance, M. Settnes {\it et al}.\cite{settnes2016valley}  and independently Milovanovi\'{c} and Peeters\cite{Peeters2016valley} proposed the use of pseudomagnetic profiles of Gaussian shapes\cite{flake2013} in order  to generate valley filtering effects, although any non-uniform strain profile is expected to exhibit such behaviour\cite{stegmann2018,Cazalilla}. Other proposals include the combination of strain effects with geometrical confinement\cite{Song,snakes-states,carrillo2016strained,milovanovic2016strained,settnes2016valley,jones2017quantized}, inclusion  of resonant structures\cite{Niu,Fujita2010,Zhenhua}, the incorporation of an artificial mass\cite{PhysRevLett.113.046601},  the addition of line defects\cite{strain-line-defect} or even under the presence of real magnetic fields\cite{Peeters2010waveTB,wu2016full,Roche} to promote valley polarization and spin-valley polarization. However, strain is not a necessary requirement to produce valley filtering effects, as local electrostatic fields alone can render the same effect as long as it is strong enough   and/or has the appropriate geometry\cite{rycerz2007valley,wang2017valley, Asmar-minimal}. 

It is also known that strain produces a scalar field that arises because of the redistribution of charge that occurs as a result of the change in the deformed area within each unit cell of graphene under elastic deformations 
\cite{AMORIM20161,suzuura2002phonons}. However most of the works on strain effects in graphene usually do not consider it\cite{Barraza01,Barraza02}. To what extent such concomitant scalar field in locally strained graphene could yield to sizable changes on the scattering phenomena is  yet a physics to be investigated. The aim of this work is to study the interplay of pseudomagnetic and scalar fields due to out-of-plane mechanical deformations in the form of Gaussian bumps in graphene and explore its role in the electron scattering. We focus our study of the electron quantum scattering problem within the Born approximation theory. The approach allow us to derive exact analytical expressions for the differential cross section for each valley $K$($K'$), treating both the pseudomagnetic and scalar fields in the same footing. Our findings  predicts that even when a relatively small scalar field is considered, a rather strong  backscattering and enhancement of the valley asymmetric scattering could arise as a consequence of its interplay with the pseudomagnetic field.  We show that the presence of the scalar field could enhance valley polarization of the scattering events as a function of the energy and angle of incidence. In addition, the valley polarization can reverse its sign  as the incident energy is increased. At first order, these behaviors are totally absent if the scalar field is neglected. In order to go beyond the Born approximation we also performed numerical simulations of the dynamics of electron wave packets and study the quantum average trajectories of the scattered wave packets. We present results of the semi-classical scattering trajectories for different angles and energies of incidence that clearly shows wave packet focusing and beam splitting effects enhanced by the presence of the scalar field.

\section{Model: graphene with a Gaussian Bump}
The dynamic of the low energy excitations in strained graphene  in the absence of interactions  is governed by the Dirac-like equation given by~\cite{sasaki2008pseudospin}
\begin{equation}\label{eq:Dirac}
i\hbar\dfrac{\partial}{\partial t}\Psi_{\eta}(\boldsymbol{r},t)  = \left[ v_{F}\boldsymbol{\sigma}_{\eta}\cdot \left(\hat{\boldsymbol{p}}-\eta{ \bm{\mathcal A} }(\boldsymbol{r})\right) +V(\boldsymbol{r}) \right] \Psi_{\eta}(\boldsymbol{r},t)
\end{equation}
where the subindex $\eta=\pm$ labels the $K$ and $K'$ Dirac points, $v_{F}$ is the Fermi velocity, $\hat{\boldsymbol{p}}=\left( \hat{p}_{x},\hat{p}_{y} \right)$ is the momentum operator of the charge carriers, and    $\boldsymbol{\sigma}_{\eta}=\left( \eta\sigma_{x}, \sigma_{y} \right)$ is the vector of the Pauli matrices. The terms  ${ \bm{\mathcal A} }$ and $V$  describe the pseudo-vector (gauge field) and the pseudo-scalar potentials, originated by the change of the carbon  bonds due to mechanical strain~\cite{sasaki2008pseudospin,neto2009electronic}. These potentials have the form
 \begin{equation}\label{eq:ScalarField}
  V=g \left( \varepsilon_{xx}+\varepsilon_{yy} \right) ,
  \end{equation}
 \begin{equation}\label{eq:GaugeField}
 { \bm{\mathcal A} }=\left( \mathcal A_{x}, \mathcal A_{y} \right)= \dfrac{\hbar\beta }{2 a_{cc}} \left( \varepsilon_{xx}- 
  \varepsilon_{yy},-2\varepsilon_{xy} 
  \right) ,
  \end{equation}
where $g$ describes the coupling with long-wave acoustical phonons due the screening with the pseudo-scalar potential in graphene, having a wide range of energy values, from 0 to 20 eV~\cite{suzuura2002phonons,vozmediano2010gauge}. The parameter   $a_{cc}=1.42 \text{ \AA}$ is the carbon-carbon interatomic distance for the unstrained graphene. The  dimensionless constant coefficient $\beta\simeq 3.0$ characterizes and tunes the effect of strain on the hopping parameter, and \(\varepsilon_{\mu\nu}\) is the strain tensor, which is defined  in terms of the in-plane displacement components \(u_{\nu}\) with $\{\mu,\nu\} = x,y$ 
and out-of-plane \(h\) deformations. The strain tensor is  dictated by the following general expression~\cite{landau1959course},
  \begin{equation}\label{eq:strainTensor}
  \varepsilon_{\mu\nu}=\dfrac{1}{2} \left(\partial_{\nu} u_{\mu}+\partial_{\mu} u_{\nu}
  +\partial_{\mu} h \partial_{\nu} h \right) .
  \end{equation}
Here we shall consider only out-of-plane deformations to model the nanoscale bump in graphene, thus Eq.(\ref{eq:strainTensor}) reduces to

  \begin{equation}\label{eq:strainTensor2}
  \varepsilon_{\mu\nu}=\dfrac{1}{2} \left(\partial_{\mu} h \partial_{\nu} h \right) .
  \end{equation}
For the analytical model of the bump itself we consider a centro-symmetrical Gaussian-shaped deformation described by the following expression
\begin{equation}\label{eq:shape}
  h(x,y)= h_o\exp{\left( -\dfrac{x^{2}+y^{2}}{b_o^{2}} \right)},
\end{equation}
\noindent where $h_o$ fixes the height of the bump,  and $b$ its effective width. The nature of the gauge field $\bm{\mathcal A}$ in Eq.\,(\ref{eq:GaugeField}) can be interpreted as a pseudo-vector potential\cite{vozmediano2010gauge} such that its  corresponding pseudo-magnetic field\cite{sasaki2008pseudospin} $\bm{\mathcal B}_{ps}$ can be written as
\begin{equation}\label{eq:FieldB}
\bm{\mathcal B}_{ps}=\eta \nabla \times \dfrac{1}{e} \bm{\mathcal A}\, ,
\end{equation}
where $e$ is the electron charge. Clearly the sign of $\bm{\mathcal B}_{ps}$ is valley-dependent and  has units of magnetic field. 
Notice that the Hamiltonian associated to Ec.(1) is symmetric under charge conjugation since the charge $q$ does not appear here explicitly in front of the pseudovector potential $\bm{\mathcal A}$.

However it does appears with opposite signs for different valleys, preserving the global time-reversal symmetry~\cite{sasaki2008pseudospin}. It is has been  already discussed that the conjugation of such symmetries can generate pseudo-spin polarization~\cite{carrillo2014gaussian,schneider2015local,georgi2017tuning}, valley splitting~\cite{stegmann2016valley,settnes2016valley} and valley filtering~\cite{Peeters2016valley,carrillo2016strained} in strained graphene.

\begin{figure}[!htbp]
\begin{center}
\includegraphics[scale=0.19]{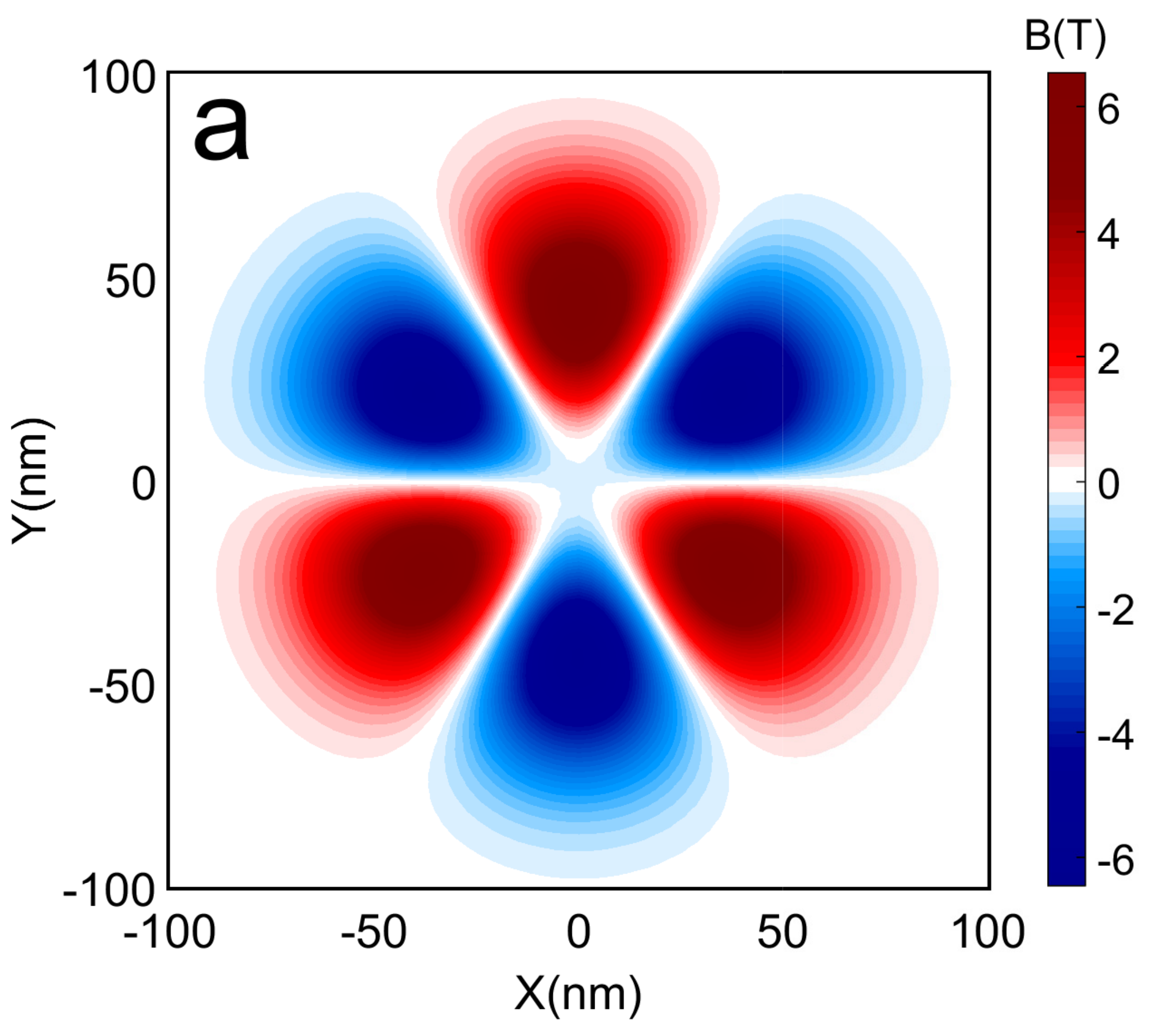}
\includegraphics[scale=0.19]{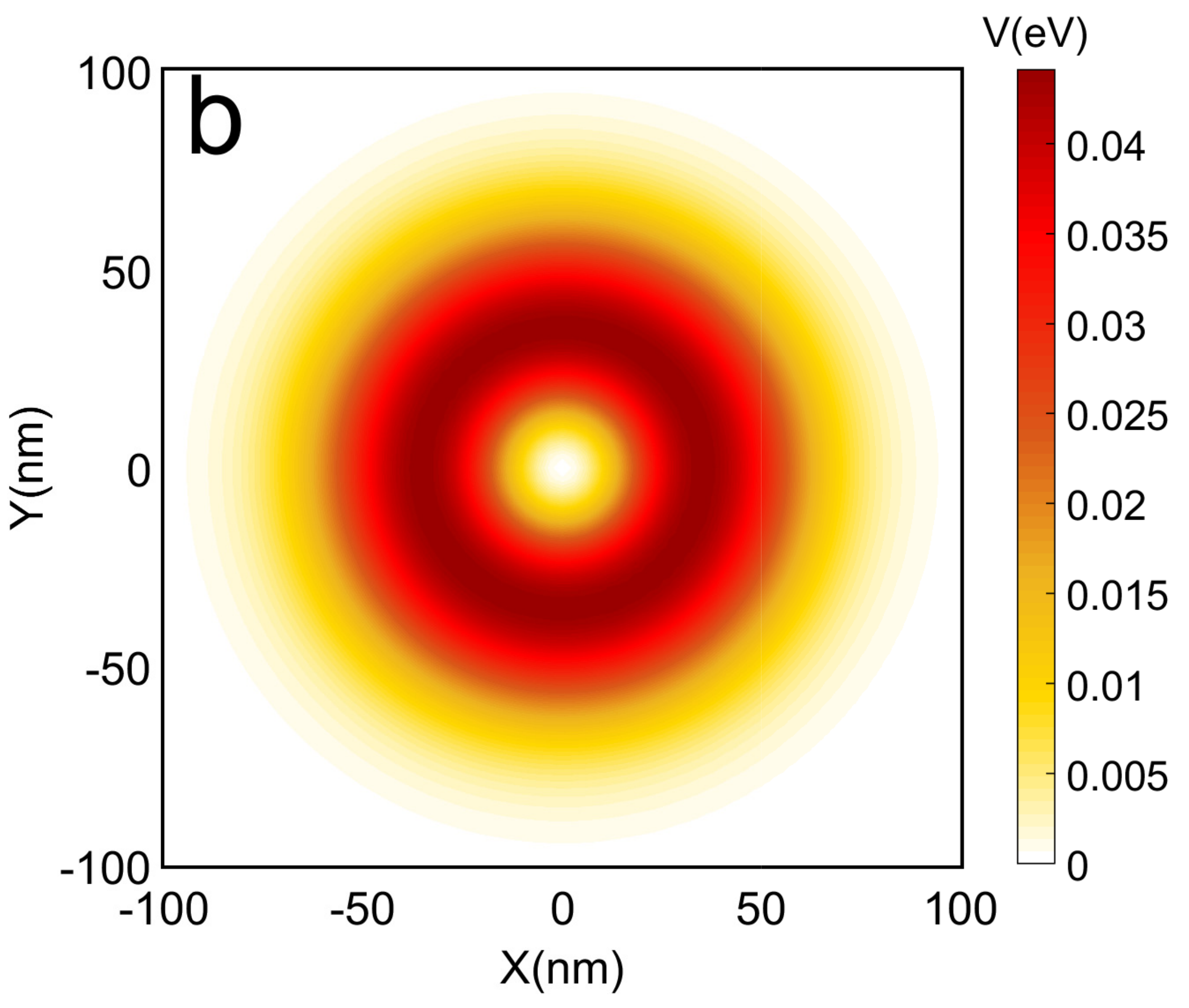}
\end{center}
\caption {(Color online) Pseudomagnetic field (a) and scalar field (b) at valley $K$ produced by a Gaussian bump of height $h_o=10\text{ nm}$, width $b_o=50\text{ nm}$, and scalar field coupling constant $g=3\text{ eV}$.
\label{fig:fields}}
\end{figure}

In this work we have considered a local Gaussian-shaped mechanical deformation in a graphene sheet with a height $h_o =10\text{ nm}$ and width $b=50\text{ nm}$. We then proceed to study the electron scattering and wave packet dynamics with ($g\neq0$) and without the presence of the scalar field  ($g=0$). For illustration, plots of the pseudomagnetic and scalar fields are shown in Fig.\ref{fig:fields} for $g=3\text{ eV}$. Notice that taking such value does not imply that the scalar field will go as high in energy, actually, for such relatively large $g$-value of the deformation, the maximum value for the scalar field achieved is just $V_{max}=0.0441\,\text{eV}$. In both, scattering and wave packet dynamics, we fix the incident energy at $E=0.11\,\text{eV}$,  and thus the corresponding incident wave number shall be given by $k_o=E/(\hbar v_F)=0.167$nm$^{-1}$.

\section{Quantum Scattering Properties}
We now look at the problem of the electron scattering phenomena in graphene with strain induced pseudomagnetic/scalar fields within the Born approximation. For this we take as weak perturbation (scattering potential) the deformation  terms in the Hamiltonian associated with Ec.\,(1) that are independent of the momentum operator, namely
\begin{equation}\label{eq:H1}
U_{\eta}(\boldsymbol{r})=-\eta v_{F}\boldsymbol{\sigma}_{\eta}\cdot\bm{\mathcal A}+V(\boldsymbol{r})\, .
\end{equation}
Due the rotational symmetry of the deformation perpendicular to the graphene sheet considered, Ec.(\ref{eq:H1}) can be rewritten in polar coordinates as
\begin{equation}\label{eq:perturbation}
U_{\eta}=\dfrac{1}{2}\left(\dfrac{\partial h }{\partial r}\right)^2 \begin{bmatrix}
    g       & -e^{2i\eta\phi}\Gamma  \\
    -e^{-2i\eta\phi} \Gamma      & g 
\end{bmatrix} 
\end{equation}
\noindent with $r=\sqrt{x^2+y^2}$, $\phi=\text{atan}(y/x)$, and   
\begin{equation}\label{eq:betaraya}
\Gamma=\dfrac{\hbar\beta}{2a_{cc}}v_{F}.
\end{equation}
Up to first order in the Born approximation, valid for the low energy limit ($kb_0\ll 1$), the scattering probability is determined by the matrix element $U^{(\eta)}_{\boldsymbol{k}_{2},\boldsymbol{k}_{1}}=\left< \boldsymbol{k}_{2},\eta |U_{\eta}| \boldsymbol{k}_{1} ,\eta\right>$, where the normalized eigenstates are given by
\begin{equation}\label{eq:statesK}
\left|\boldsymbol{k},\eta\right>=\dfrac{1}{\sqrt{2}}\begin{bmatrix} 
e^{-i\eta\frac{\theta}{2}} \\ \pm \eta e^{+i\eta\frac{\theta}{2}} \end{bmatrix}  e^{i\boldsymbol{k}\cdot\boldsymbol{r}},
\end{equation}
with $|\boldsymbol{k}_{1}|=|\boldsymbol{k}_{2}|=k$ to ensure energy conservation during the scattering process. Thus the differential cross section per Dirac point $\eta$ in terms of the scattering probability is determinate by 
\begin{equation}
\sigma^{\eta}_{D}=\dfrac{k}{2\pi \hbar^2 v_{F}^2}|U^{(\eta)}_{\boldsymbol{k}_{2},\boldsymbol{k}_{1}}|^2.
\end{equation}
Similarly as done for the case without scalar fields\cite{yang2012scattering}, we can write the  differential cross sections as follows,
\begin{equation}
\sigma^{\eta}_{D}=\dfrac{k}{2\pi \hbar^2 v_{F}^2} \left|g\cos{(\theta_{m})}F_{k0}(\theta_{m})\mp\eta\Gamma\cos{(3\theta_{p})}F_{k2}(\theta_{m})\right|^2
\end{equation}
where we have defined the function
\begin{equation}\label{eq:Fn}
F_{kn}(\theta_{m})=\pi\int^{\infty}_{0}J_{n}\left[ 2kr\sin{\theta_{m}} \right]\left( \dfrac{\partial h}{\partial r}\right)^2rdr, 
\end{equation}

\noindent here $J_{n}(z)$ is  the Bessel function of order $n=0,2$, with $\theta_{m/p}=(\theta_{2}\mp\theta_{1})/2$, being $\theta_{1}$ the angle of incidence, $\theta_{2}$ the scattered angle, with a deformation out-of-plane  characterized by Eq.(\ref{eq:shape}) as,  
\begin{equation}
 \dfrac{\partial h}{\partial r} =-2\dfrac{h_o r}{b_o^2}\exp{\left[-\dfrac{r^2}{b_o^2}\right]}.
\end{equation}
The terms out-of-diagonal in Eq.(\ref{eq:perturbation}) generate a non-uniform  pseudo-magnetic field  with a three-fold symmetry per valley given by
\begin{equation}
\bm{\mathcal B}_{ps}=\eta \dfrac{2h_{o}^2 B_{o}}{b_o^2} \left( \dfrac{r}{b_o}\right)^3 e^{-2(r/b_o)^2}\sin(3\phi) \hat{z} ,
\end{equation}
with $B_{o}=4\Gamma/(ev_{F}b_o)$, whereas the diagonal terms act as the scalar field 
\begin{equation}\label{eq:ScalarExp}
V=2g\left( \dfrac{r}{b_o}\right)^2e^{2(r/b_o)^2}.
\end{equation}
The radial integrals defined in Eq.(\ref{eq:Fn}) can be obtained analytically, having the following closed form (see Appendix, section \ref{sec:F2y0}),
\begin{equation}\label{eq:F0}
F_{k0}(\theta_{m})=\dfrac{\pi h_o^2 }{2}[1-\lambda^2_{k}]e^{-\lambda^2_{k}},
\end{equation}
\begin{equation}\label{eq:F2}
F_{k2}(\theta_{m})=\dfrac{\pi h_o^2 }{2} \lambda^2_{k}e^{-\lambda^2_{k}},
\end{equation}
\noindent with $\lambda^2_{k}=k^2b_o^2\sin^2(\theta_m)/2$. Although previous scattering studies\cite{yang2012scattering,munoz2017,liu-low} have considered similar effects of the pseudomagnetic field as studied here, however, the effect of the concomitant scalar field  itself was ignored in these works. Moreover, while in Ref.[\onlinecite{yang2012scattering}]  only provide of approximate expressions for the radial integral involved in the calculation of the differential cross section, here in contrast, we were able to provide exact  analytical formulas for these integrals even for the case of the presence of the scalar field effect. After evaluating the integrals, we can write the exact differential cross section as,
\begin{equation}\label{eq:sigmaD}
\sigma^{\eta}_{D}=\dfrac{\pi k h_o^4}{ 8\hbar^2  v_{F}^2} \left|g\cos{(\theta_{m})}(1-\lambda^2_{k})\mp\eta\Gamma\cos{(3\theta_{p})}\lambda^2_{k}\right|^2 e^{-2\lambda^2_{k}}.
\end{equation}
This is one of the main results of this work.
Clearly at first order and in the absence of the scalar field ($g=0$), there is no difference between the contribution to the quantum scattering of the  $K$ and $K'$ Dirac points to the total differential  cross section ($\sigma^{+}_{D}=\sigma^{-}_{D}$). However, for $g\ne0$ we have that the differential scattering cross sections $\sigma^{+}_{D}\neq\sigma^{-}_{D}$ in general, which points out the importance of considering on the same footing the interaction of both fields, as we shall discuss in more detail below for specific cases.

\begin{figure}[!htbp]
\begin{center}
\includegraphics[scale=0.20]{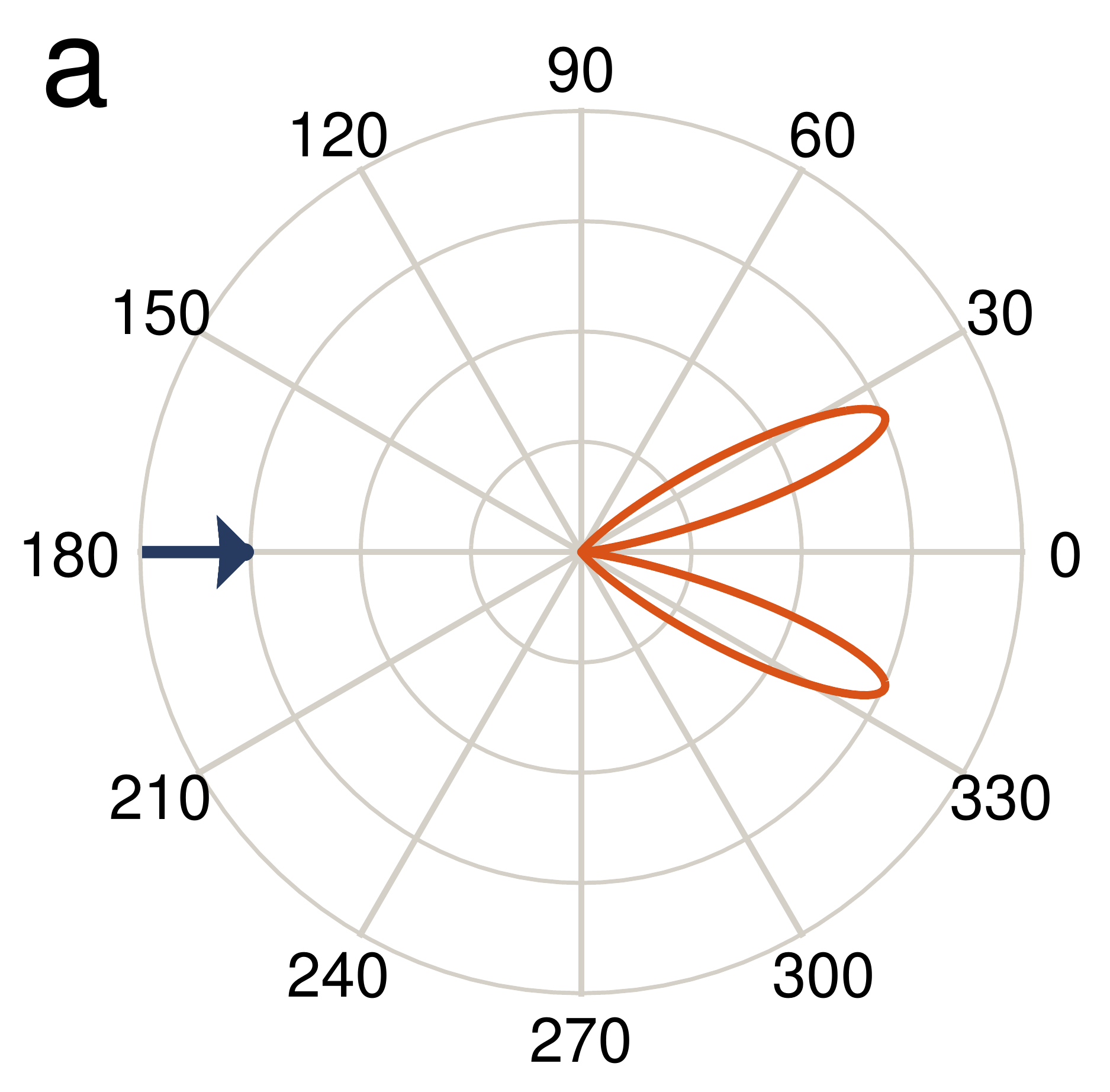}
\includegraphics[scale=0.20]{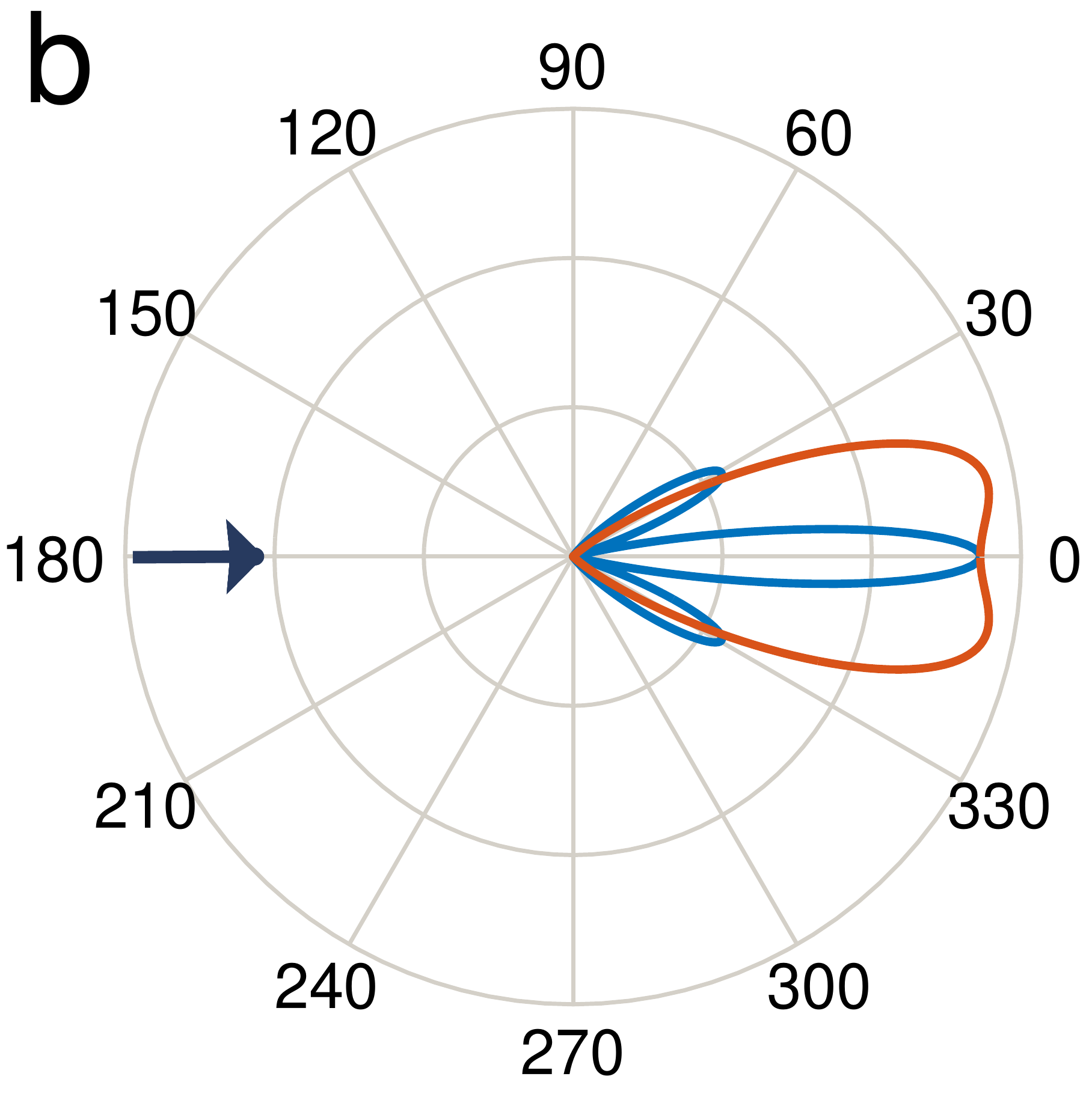}
\includegraphics[scale=0.20]{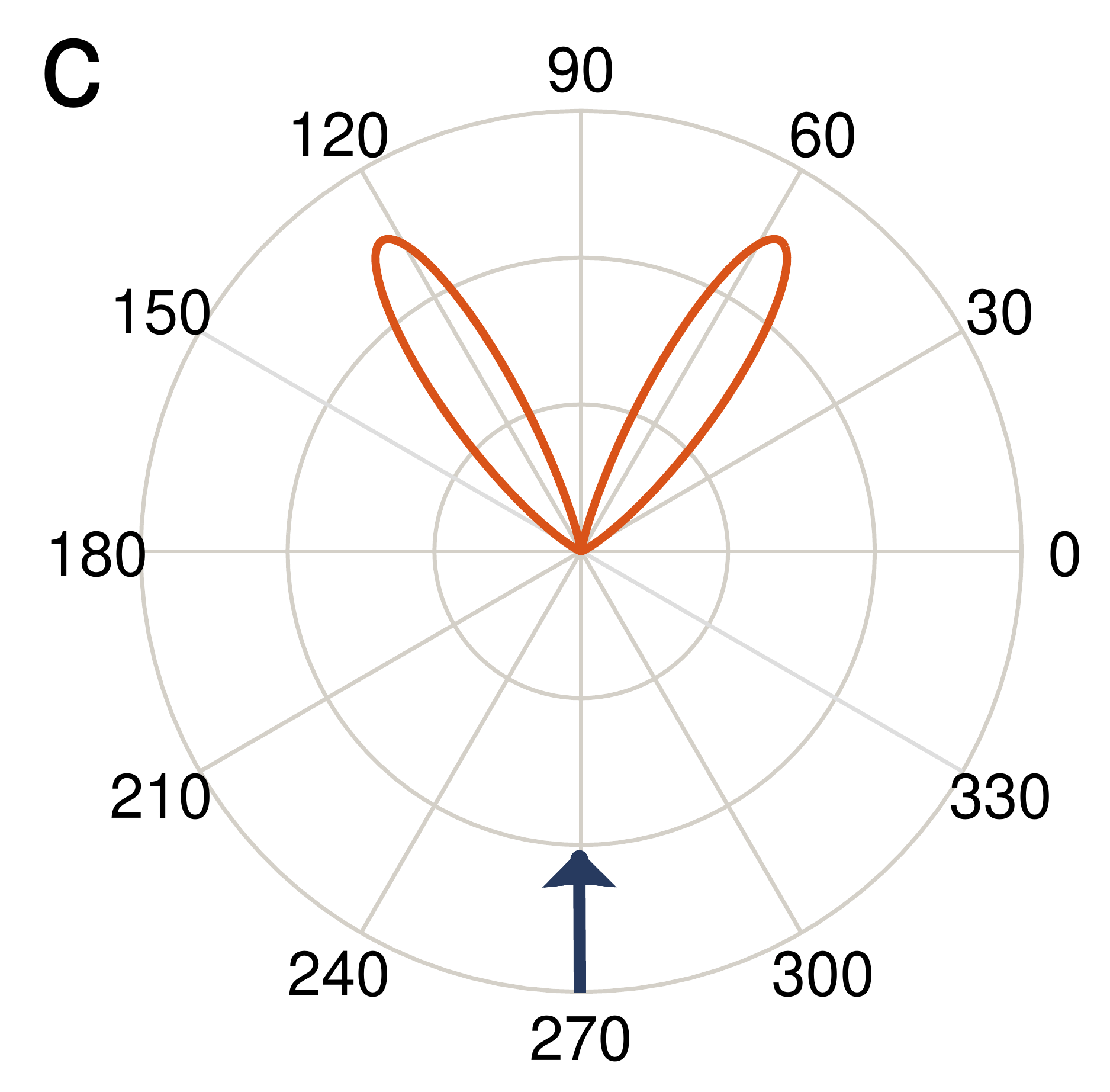}
\includegraphics[scale=0.20]{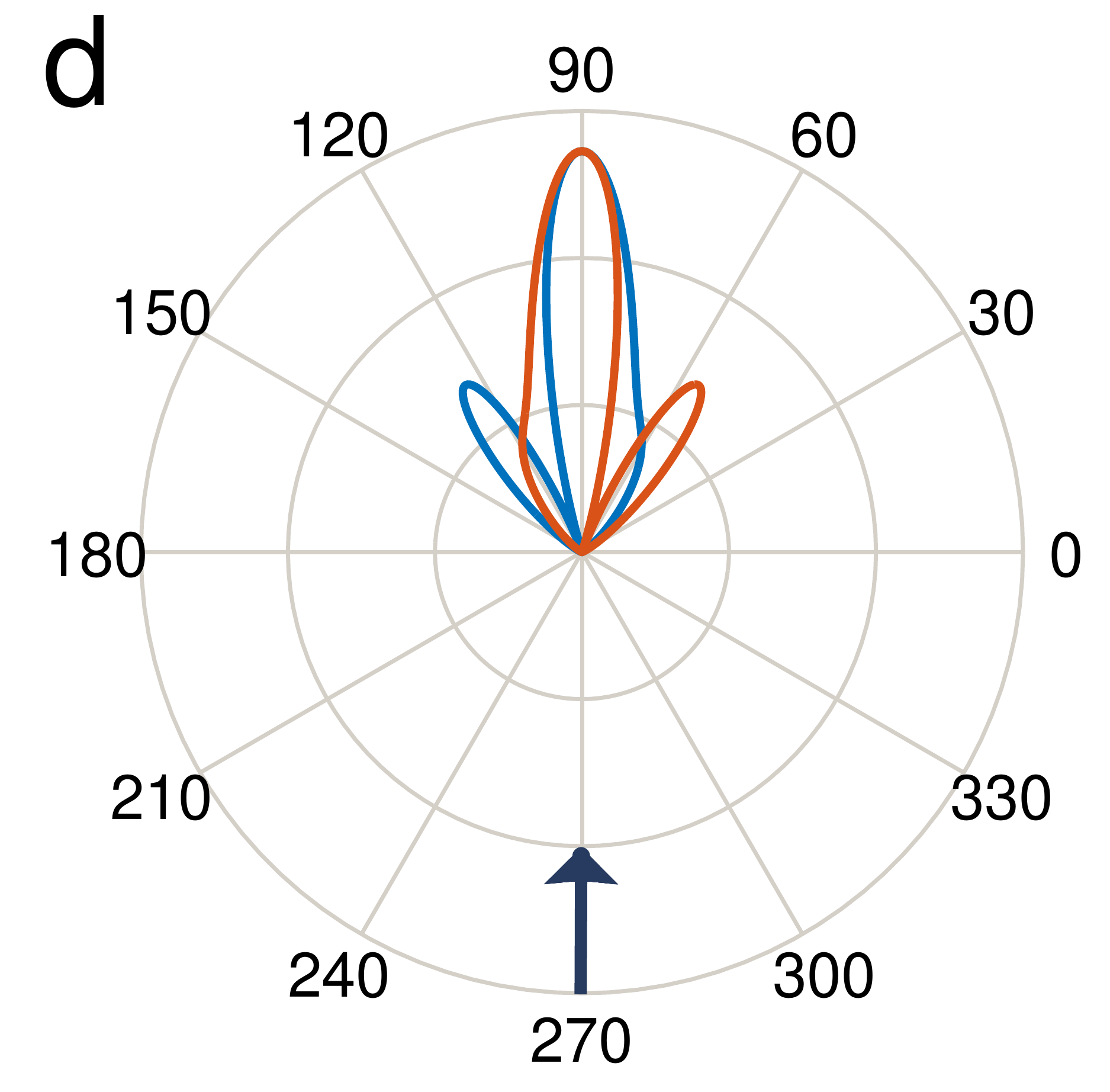}
\end{center}
\caption {(Color online) Differential scattering cross section $\sigma_D^{\pm}$ for valley $K$ (blue) and $(K')$ (orange) as function of the outgoing angle $\theta_2$ for horizontal incidence (first row) and vertical incidence (second row); in the left column $g=0$ and  for the right column $g=3\text{ eV}$.
\label{fig:sd}}
\end{figure}

In Fig.\ref{fig:sd} we depict in polar plots the differential scattering cross section $\sigma_D^{\eta}$  as a function of the angle $\theta_2$ of the out-going (scatter) wave for horizontal and vertical incidence, respectively.  The incident direction is shown by a dark blue arrow with an incident effective momentum of $k=0.167\text{ nm}^{-1}$. Blue curves represent the contribution for $K$-point (valley) to the differential scattering cross section, while orange curves are from $K'$-point(valley) contribution. In the left panels (a and c), the scalar field is set to zero, ($g=0$), therefore the curves for each $\eta=\pm$ overlaps as expected from Eq.(\ref{eq:sigmaD}). On the other hand, in the right panels (b and d) we are taking $g=3\text{ eV}$. Notice that for horizontal incidence, the interplay of the scalar and pseudomagnetic fields promotes the appearance of a narrow angular region with rather different angular distribution of the scattering cross section of the $K,K'$ valleys (see e.g. Fig.~\ref{fig:sd}b).   
In fact, such horizontal incidence configuration has been proposed earlier in the literature\cite{Peeters2016valley,settnes2016valley,georgi2017tuning} as a possible valley splitter for vanishing $g$. Here we find that vertical incidence can also generate narrow angular distribution of the scattering cross section that can give rise to sizable valley polarization at $g=3$ eV, as shown in Fig.~\ref{fig:sd}d.
\begin{figure}[!htbp]
\begin{center}
\includegraphics[scale=0.22]{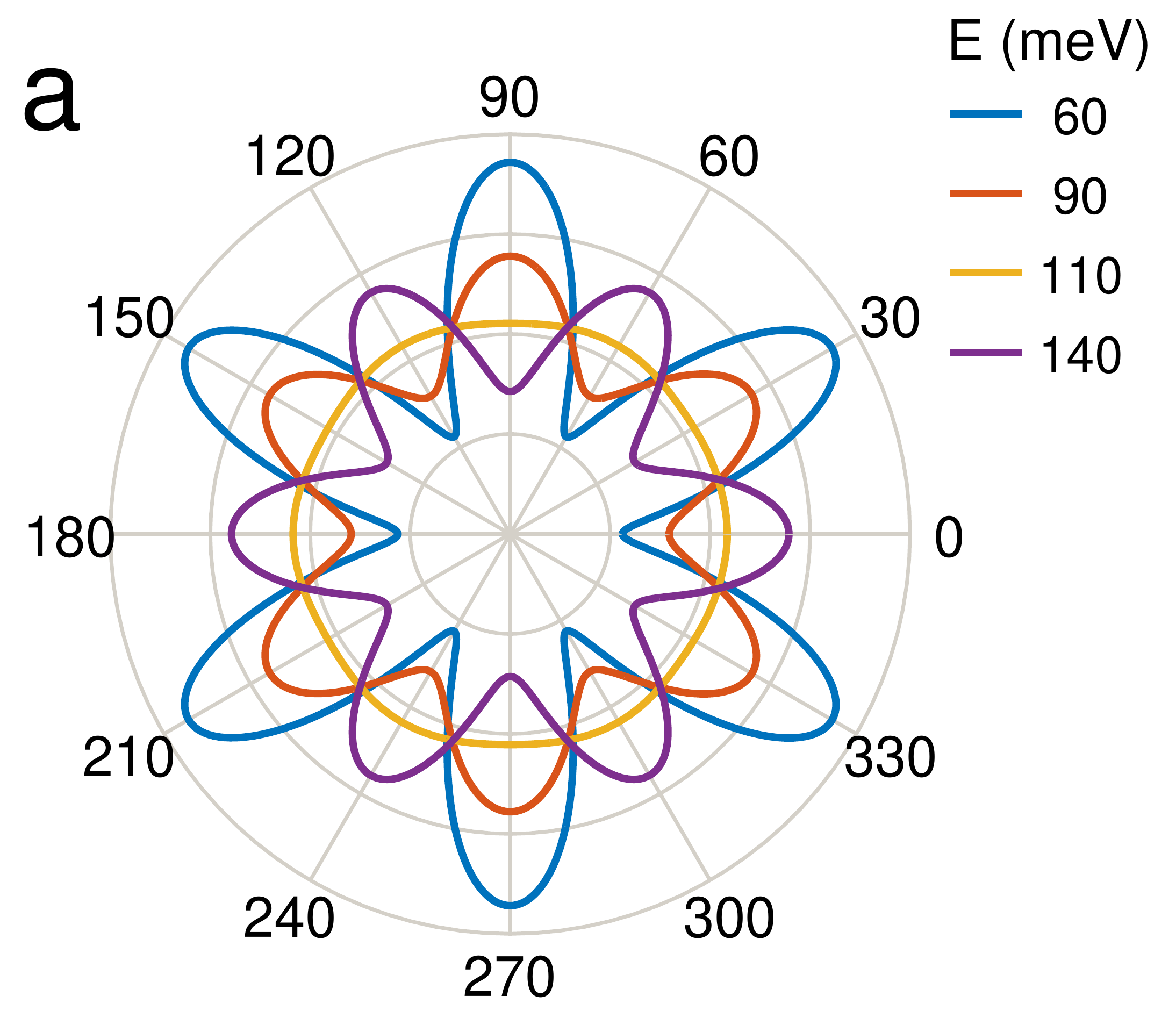}
\includegraphics[scale=0.22]{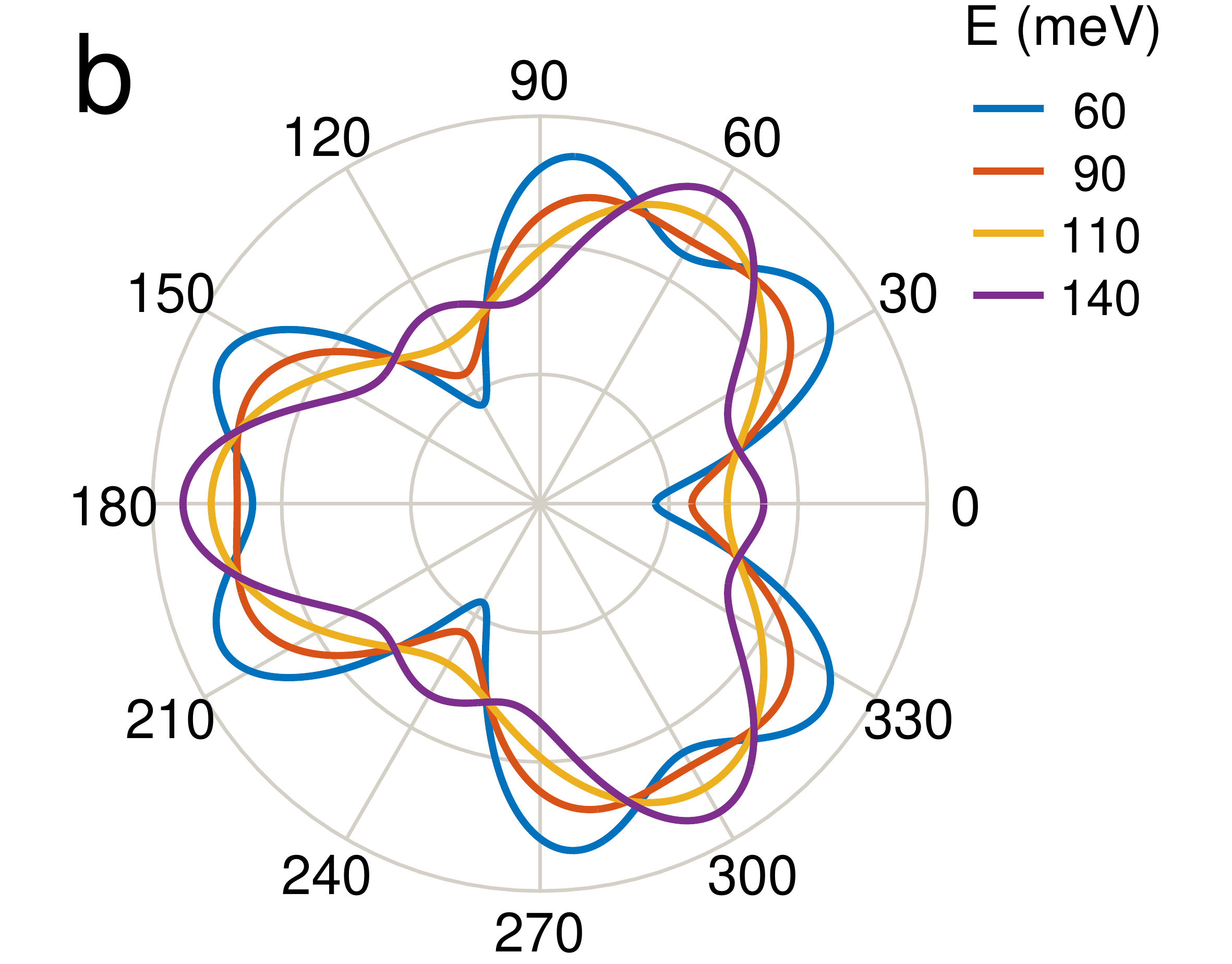}
\includegraphics[scale=0.22]{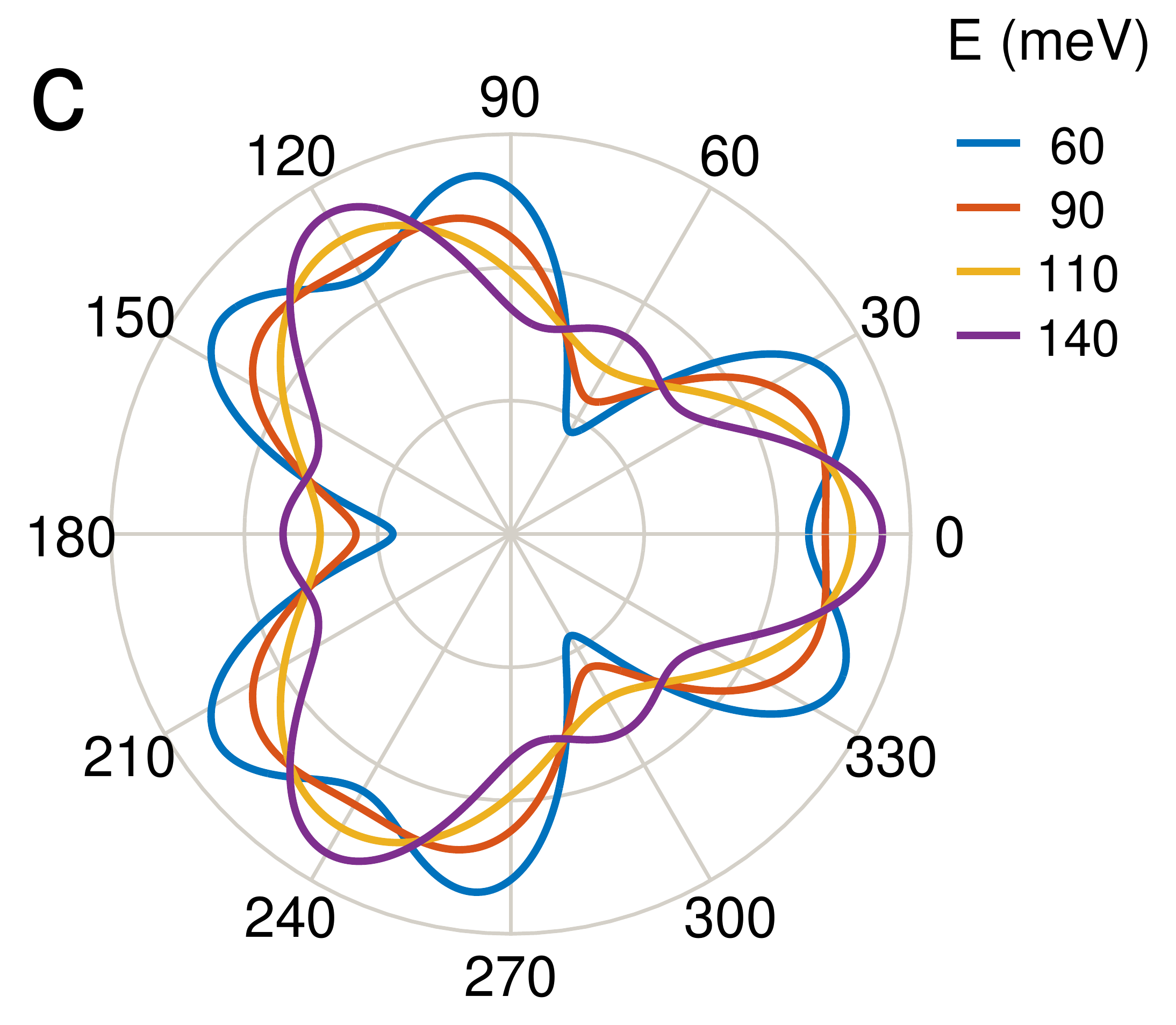}
\end{center}
\caption {(Color online) Total scattering cross section $\sigma_T$ as function of the incident angle $\theta_1$ for (a) $g=0$, and $g=3$~eV for valley $K$ (b) and $K'$ (c).
\label{fig:sT}}
\end{figure}

The total scattering cross section per valley $\eta$ is given by
\begin{equation}
\sigma^{\eta}_T=\int_{0}^{2\pi} \sigma^{\eta}_{ D}\text{\small d}\theta_2.
\end{equation}
Plots of the total scattering cross section as a function of the incident angle $\theta_1$ for different energies from $60$ to $140$\, meV are shown in Fig.~\ref{fig:sT}. In panel (a), we are ignoring the scalar field ($g=0$), and therefore the results are identical for both $K,K'$ valleys. The  behavior $\sigma^{\eta}_{D}$ with energy is non monotonic, instead it oscillates with the incident energy (below we discuss this dependence), and for certain values of energy it shows an uniform angular distribution. In panels (b) and (c), we present $\sigma^{\eta}_T$  as a function of the incident angle, for $g=3\text{ eV}$. 
\begin{figure}[!htbp]
\begin{center}
\includegraphics[scale=0.205]{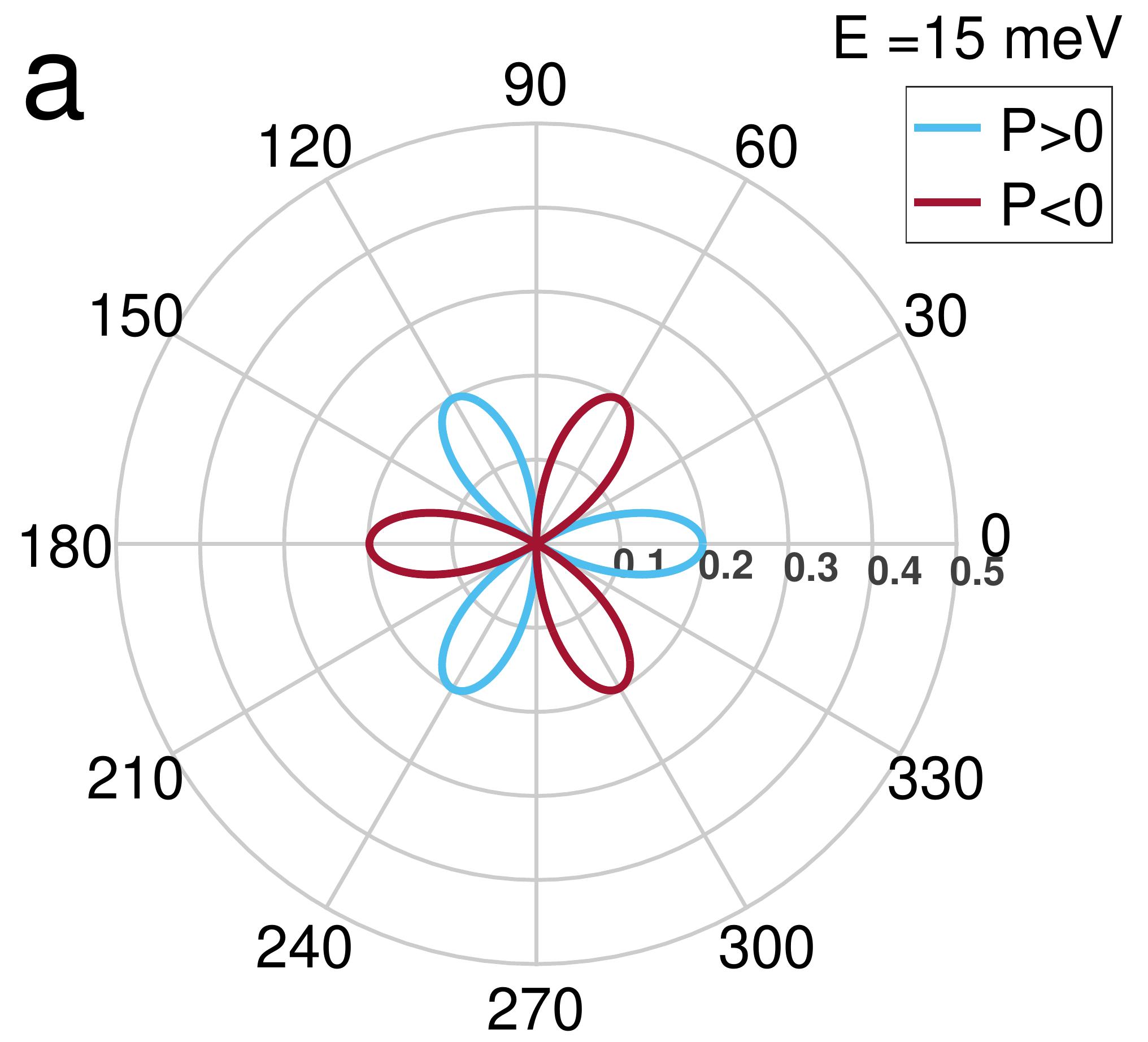}
\includegraphics[scale=0.205]{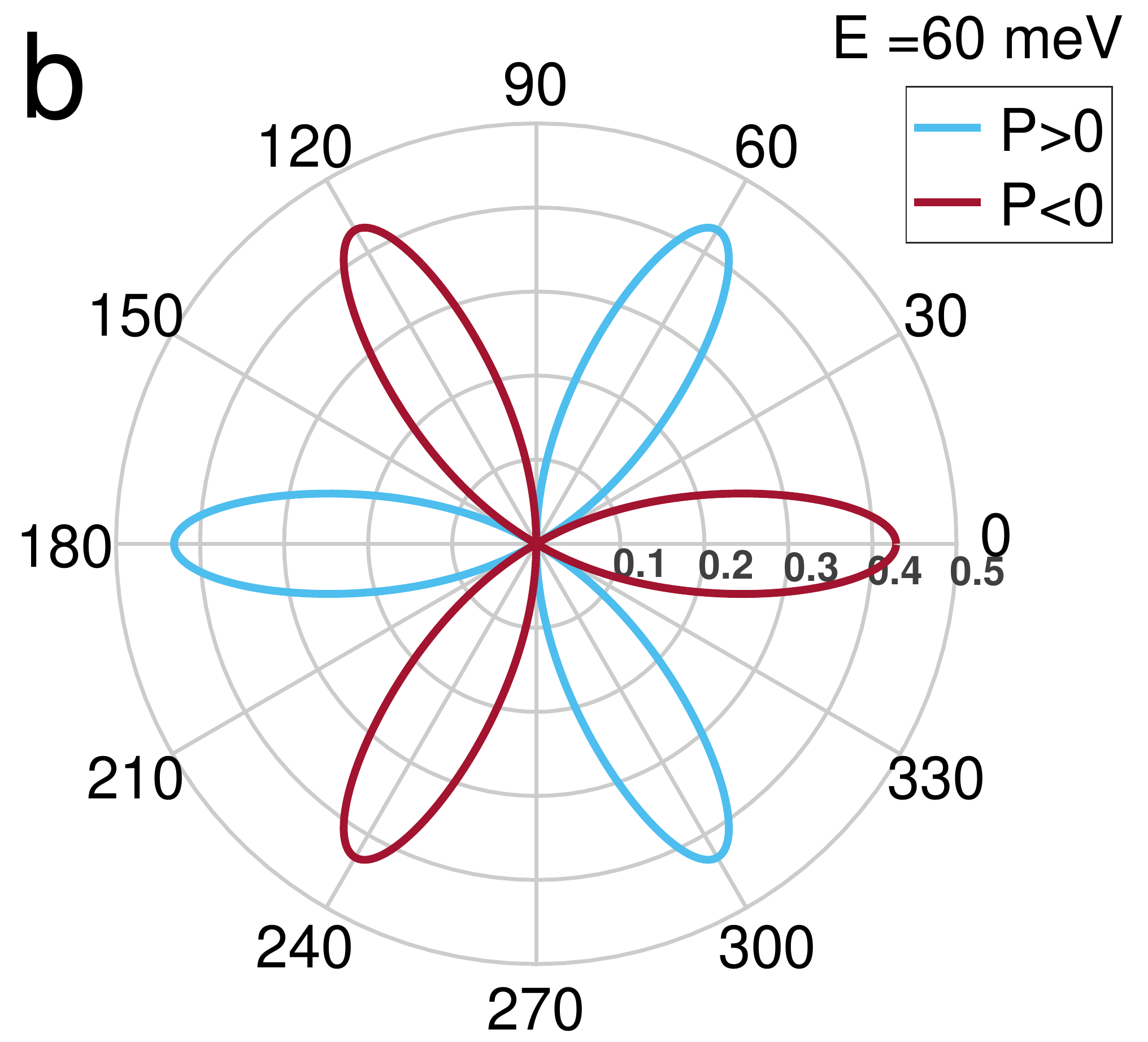}
\includegraphics[scale=0.29]{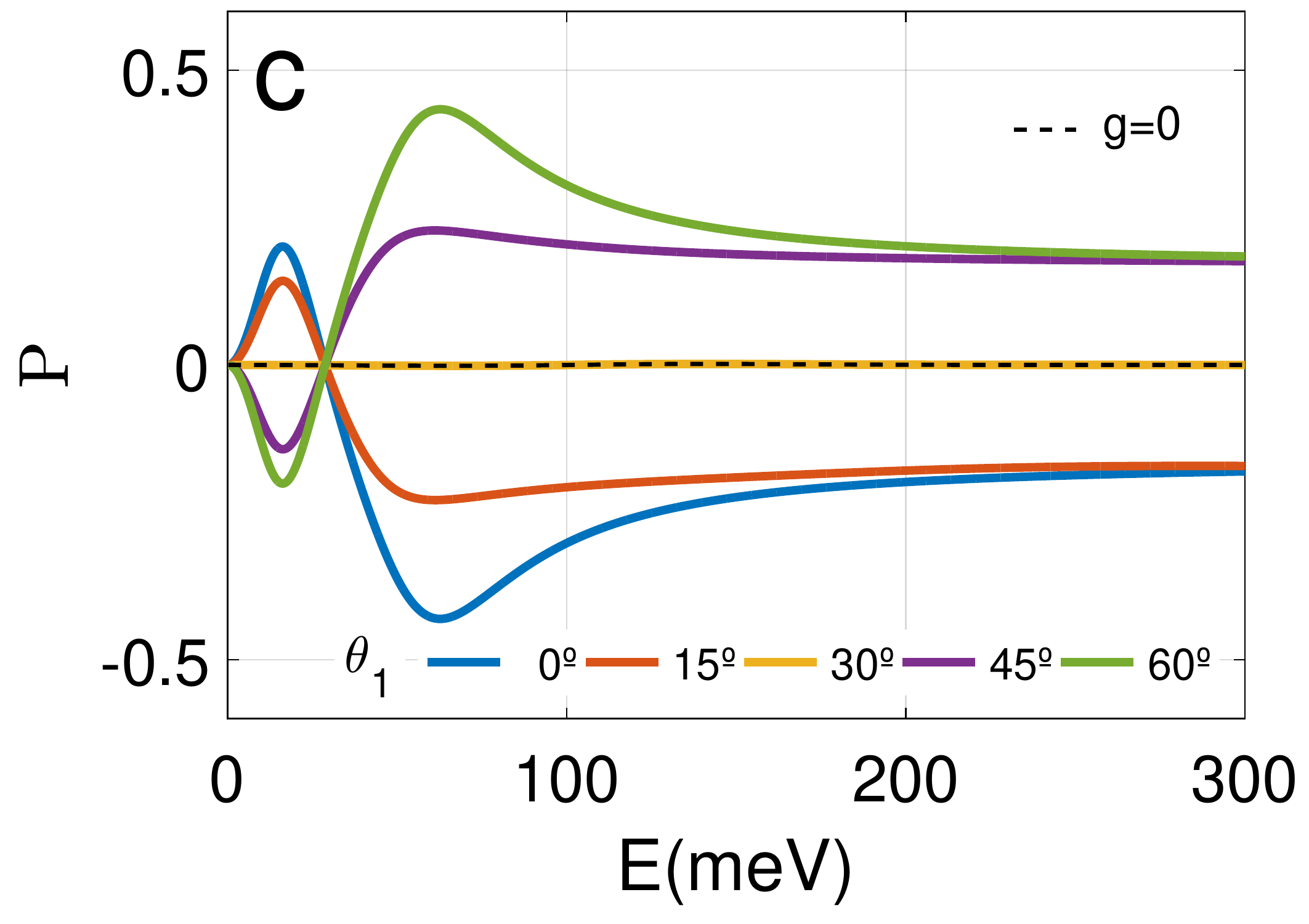}
\end{center}
\caption {(Color online) Valley polarization efficiency ${\cal P}$ as function of the incident angle $\theta_1$ for incident energy $E=15$ meV (a) and $E=60$ meV (b). Panel (c) shows the ${\cal P}$ as function of the incident energy for  different incident angles $\theta_1$. We take $g=3$~eV, except in the dashed line where $g=0$~eV.\label{fig:pol}}
\end{figure}
Though there is an oscillating behavior, the plots clearly shows that for certain angles the scattering for one valley is maximum, while for the other valley is minimum. For example at $60^\circ$ angle, the $\sigma^{\eta}_T$ is a maximum for valley $K$ and a minimum for valley $K'$, and the situation reverses at the angle of $120^\circ$. Thus a valley splitting effect is expected at these particular angles of dispersion. The situation will be the same for horizontal incidence and it will repeat each $60^\circ$ angle because of the six-fold symmetry of the pseudomagnetic field. Notice that for vertical incidence {\it i.e. }$\theta_1=270^\circ$, yields $\sigma^{\eta}_T$ identical for both valleys (Fig.~\ref{fig:sT}b and Fig.~\ref{fig:sT}c). Nevertheless as we discuss above, the direction of scattering will differ for each valley (see Fig.~\ref{fig:sd}d). 

To characterize further the valley polarization, we define a valley polarization efficiency ${\cal P}$, as  
\begin{equation}
{\cal P}= \frac{\sigma^{+}_T-\sigma^{-}_T}{\sigma^{+}_T+\sigma^{-}_T},
\label{eq:polarization}
\end{equation}
which is positive for the $K$ valley polarization and negative for $K'$. We show several plots for relevant cases in Fig. \ref{fig:pol}. In panel  a)  we fix the incident energy to $E=15$~meV and plot $\cal{P}$ as function of the incident angle $\theta_1$. Panel b) Fig.~\ref{fig:pol} shows the corresponding results for $E=60$~meV. Both plots show a six-folded (three-fold per valley) structure similar to the one of the pseudomagnetic field, but with a $30^\circ$  rotation. These results support the proposal of three-terminal structures like the one in Ref.~[\onlinecite{daiara3}]. Interestingly the sign of the polarization is inverted between these two plots. The reason behind this fact is that for low energy, the back-scattering becomes dominant and become strongly K-valley dependent. The later effect can be seeing in panel c), where we plot the dependence of $\cal{P}$ with the incident energy. The dashed line correspond to the zero polarization output for    the absence of scalar field, and  the continuous  curves show results for  different incident angle $\theta_1$ with $g=3\text{ eV}$. In particular, the blue curve shows  the valley polarization efficiency $\cal{P}$, for horizontal incidence ($\theta_1=0^\circ$).  It increases with energy, presenting a maximum around $18$~meV, then it decreases uniformly till it change sign with a minimum at $60$~meV. A similar behavior is shown by the red curve, at $\theta_1=15\circ$ while the sign is inverted for the purple and green curves at $\theta_1=60^\circ$ and $\theta_1=45^\circ$, respectively. For $\theta_1=30^\circ$  the polarization is zero for all energies. All the curves show an approximated constant behavior for energies greater than $150$~meV and  with exception of the yellow one they present values of polarization close to $20\%$. We attribute the oscillatory behavior with energy to the quantum backscattering effect that becomes relevant for small energies ($i.e. ~\lambda_o \sim b_o$).

It has been shown using Boltzmann transport equation that the electronic transport in graphene under strains is mainly governed by
the acoustic gauge field, while the contribution due to the deformation potential may be negligible and strongly screened\cite{sohier2014phonon}. Clearly in such cases the valley-asymmetric scattering showed in Eq. (\ref{eq:sigmaD}) will not be present. However is very important to remark that a similar valley-asymmetric scattering behavior is expected in presence of any other scalar potential even if they are not produced by strain. Consider for example a scalar potential proportional to the height of the membrane $h(x,y)$, namely
\begin{equation}
V_{ext}=\epsilon\mathcal{A}\exp{\left( -\dfrac{x^{2}+y^{2}}{b_o^{2}} \right)}=\epsilon h(x,y).
\end{equation}
\noindent being $\epsilon$ some constant. Such a field will appear for instance if there is a nonuniform electric field pointing to the $z$-direction perpendicular to the membrane , as the case of AFM-tip \cite{georgi2017tuning} and/or a gate\cite{klimov} pulling the membrane. More recently, this shape of potential has been used to model a screened confining potential caused by ionized impurities \cite{Wedding2018}. Taking $g=0$ (ignoring the scalar field associated with strain) the corresponding differential cross section is instead,
\begin{equation}\label{eq:sigmaD-s}
\sigma_{D}^{\eta}=\dfrac{\pi k h_o^2}{ 8\hbar^2  v_{F}^2} \left|2\epsilon\cos{(\theta_{m})}e^{-\lambda^2_{k}}\mp\eta\widetilde{\beta}\cos{(3\theta_{p})}\lambda^2_{k}\right|^2 e^{-2\lambda^2_{k}}.
\end{equation}
Hence, a non-uniform electric field will also generate valley polarization and similar angular dependence as in Eq.~(\ref{eq:sigmaD}) even in the absence of the scalar field produced by the strain.

It is worthwhile to emphasize that the physics discussed in this section refers strictly to the first order Born approximation results, and as such some relevant phenomena can be ignored. As a matter of fact, it is well known that the pseudomagnetic field produced by centrosymmetrical deformations generates by itself a valley-asymmetric scattering\cite{georgi2017tuning,settnes2016valley,Peeters2016valley}. However, as shown here, in the absence of a scalar field, the expected valley asymmetric scattering cannot be obtained within the first order Born approximation. As pointed out in Refs.\cite{georgi2017tuning,zhai2018local}, it is necessary to go up to second order in the Born expansion  for the valley asymmetric scattering to appear.   Nevertheless, the scalar field can modify the valley asymmetric splitting. In particular, it makes the valley polarization present even at first order, contrary with the case without scalar field. On the other hand, the presence an external scalar field scenario (by gating for instance), opens the possibility of controlling the valley scattering polarization by electrical means.

\section{Wave packet Propagation}

The Born approximation describes correctly the scattering in the limits of low and high energies. In order to go beyond the Born approximation and explore the intermediate energy regime, we study the dynamics of the scattering process of electron wave packets in strained graphene by numerically solving  Eq.\,(\ref{eq:Dirac}) in finite differences in real space. The scheme employs  a suitable  splitting of the time evolution operator. The resulting differential equations are solved in a recursive approach for any given time step provided the initial and boundary conditions of the strained graphene sheet (for details see Appendix, section B). 

As for the initial condition, we take an incident Gaussian wave packet of standard deviation, $w$, mean position, $\boldsymbol{r}_0=(x_0,y_0)$, moving with an average momentum, ${\boldsymbol{p}}_{o}=\hbar\boldsymbol{k}_{o}$, given by,

\begin{equation}\label{eq:statesK}
\Psi^\eta _{\boldsymbol{k}_{o}}(\boldsymbol{r},0)=\dfrac{1}{\sqrt{4\pi w^2}}\exp{\left[\dfrac{(\boldsymbol{r}-\boldsymbol{r}_0)^2}{2w^2}+i\boldsymbol{k_o}\cdot\boldsymbol{r}\right]}\begin{bmatrix} 
e^{-i\eta\frac{\theta}{2}} \\ \eta e^{+i\eta\frac{\theta}{2}} \end{bmatrix}.
\end{equation}

\noindent where $\frac{1}{\sqrt{2}}(e^{-i\eta\frac{\theta}{2}}  \,\,\, e^{+i\eta\frac{\theta}{2}})^T$ is the initial pseudospin polarization, aligned with the direction of average momentum, ${\boldsymbol{p}}_{o}$. In particular, for all our numerical simulations we use a wave packet with $w=30\text{ nm}$ and total momentum $k=0.167\text{ nm}^{-1}$, corresponding to an incident energy $E=0.11\,\text{eV}$. 
\begin{figure}[!htbp]
\begin{center}
\includegraphics[scale=0.186]{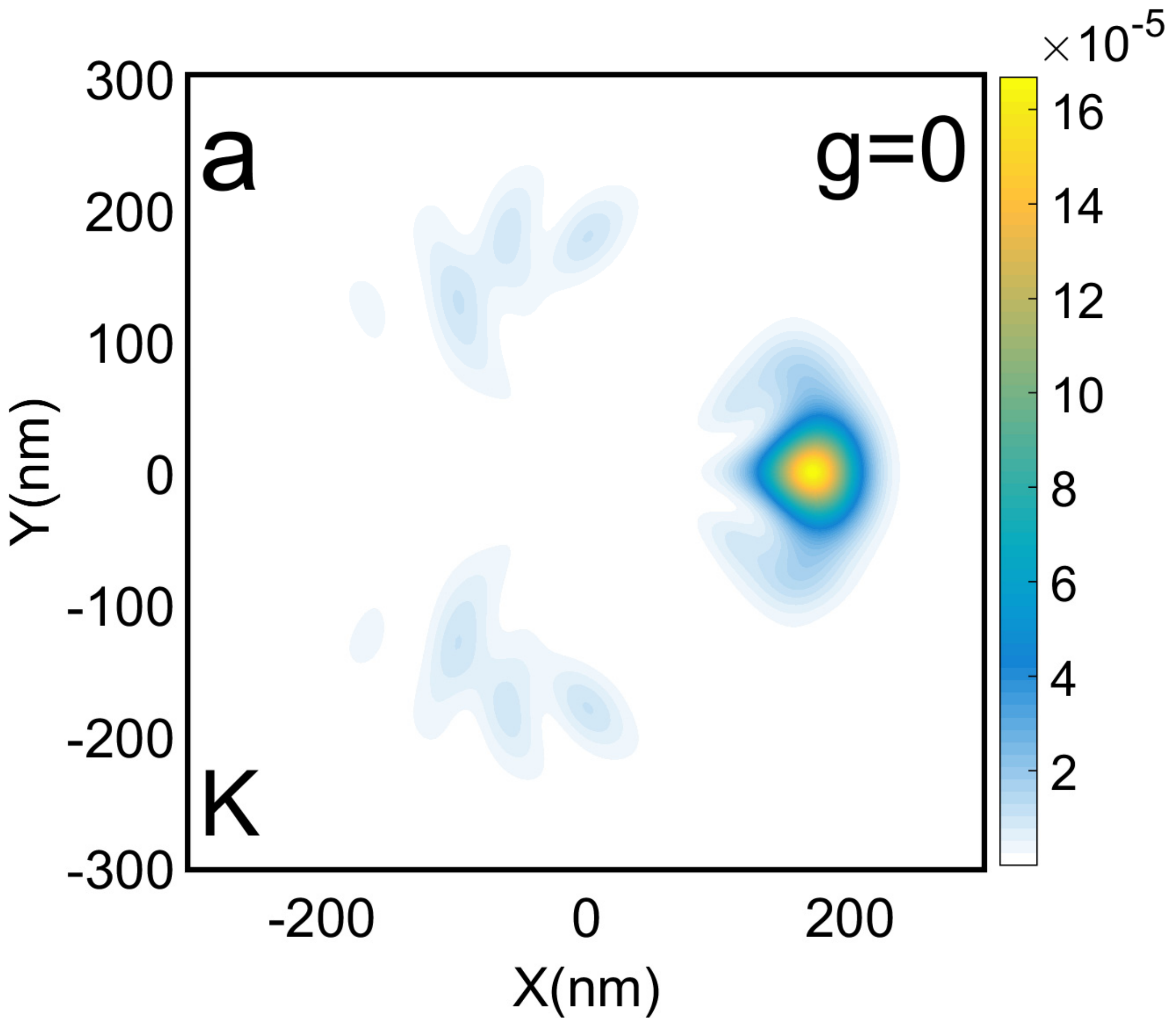}
\includegraphics[scale=0.186]{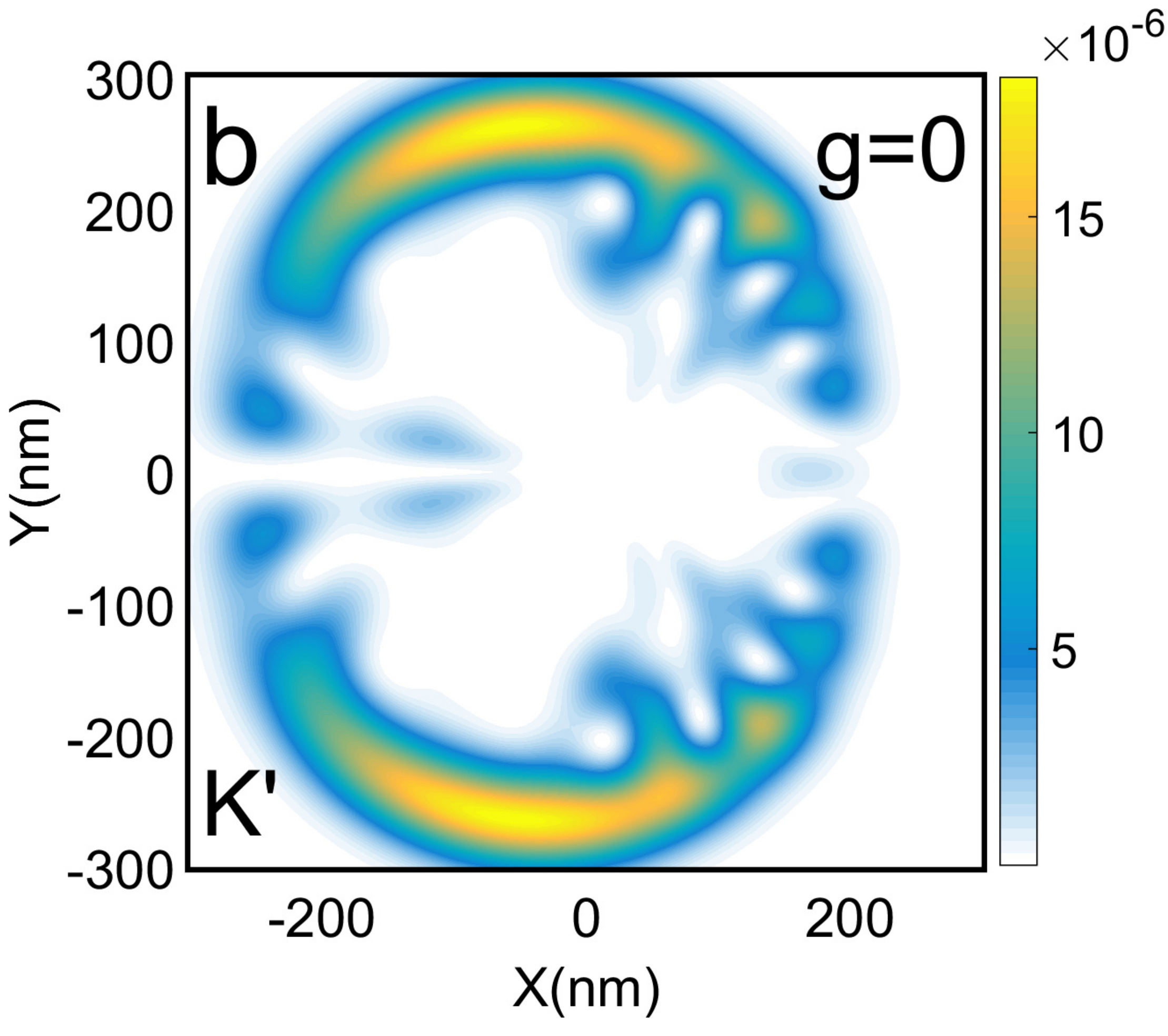}
\includegraphics[scale=0.186]{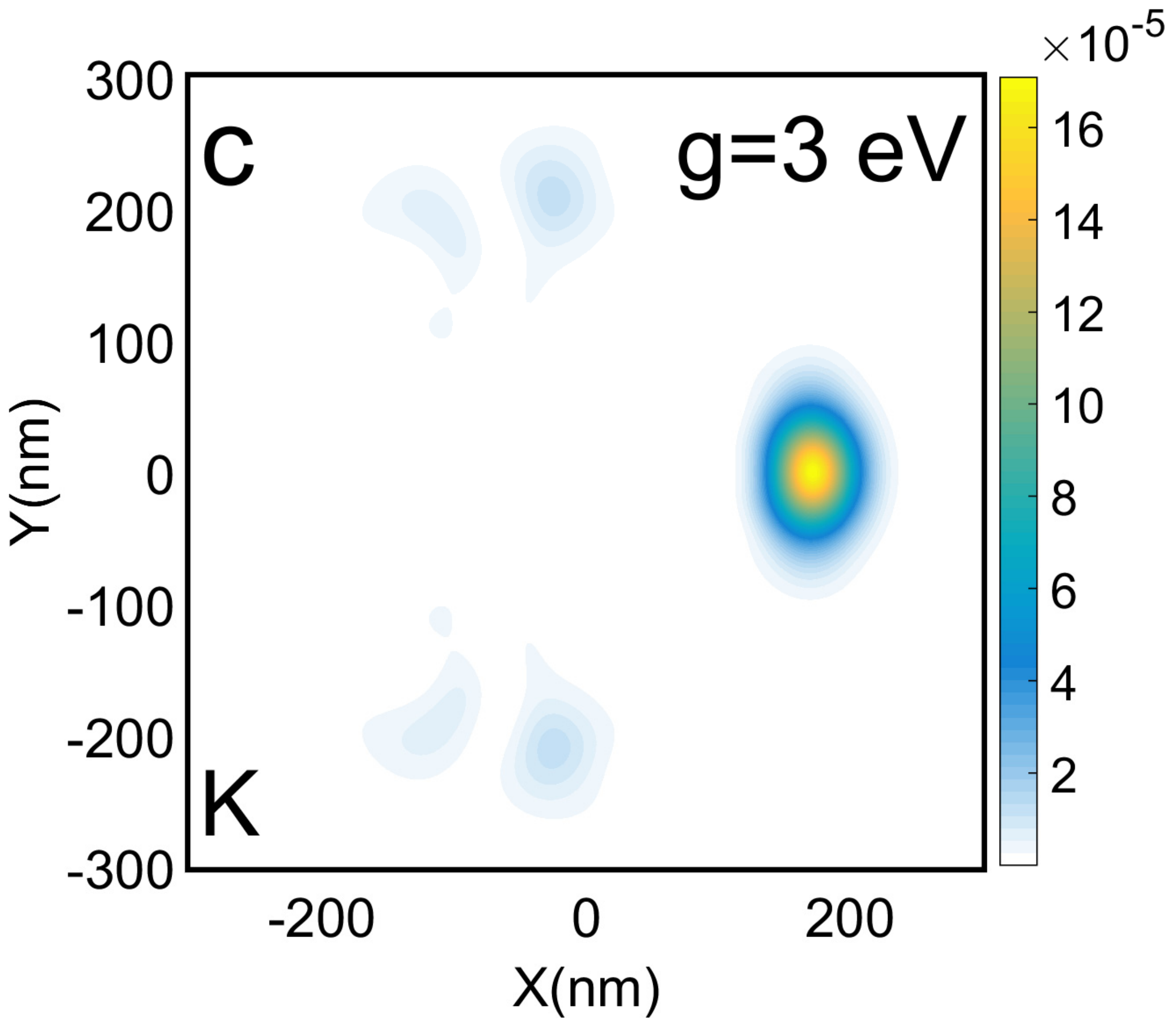}
\includegraphics[scale=0.186]{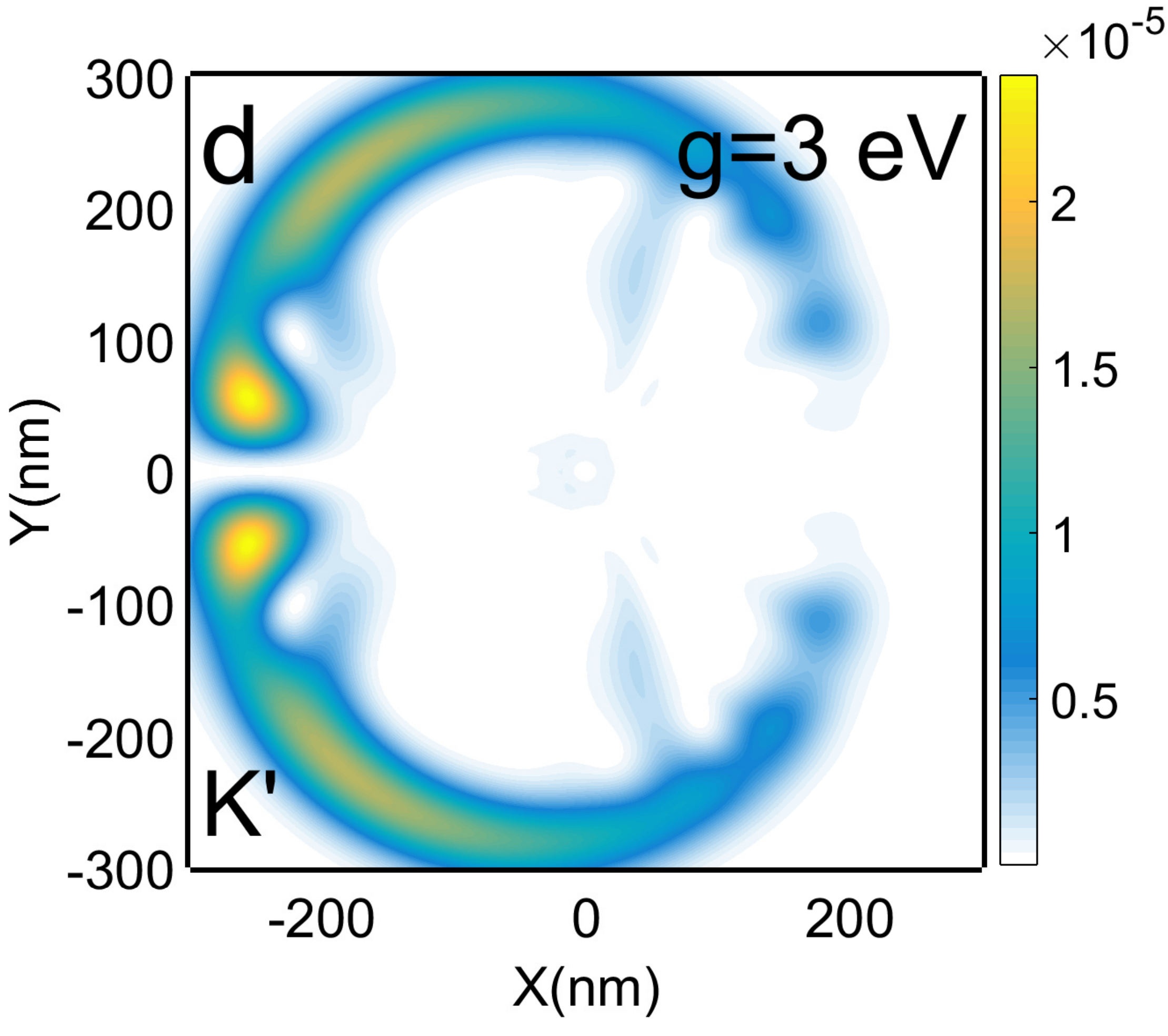}
\end{center}
\caption {(Color online) Probability density $|\Psi (x,y)|^2$ of an incident wave packet coming from the left [$(x_0,y_0)=(-150,0)\text{ nm}$] with $E=110\text{ meV}$ at time, $t=350\text{ fs}$  with $g=0$ (first row) and $g=3$~eV (second row). Different columns correspond to different valleys: (a) and (c)  to  valley $K$, and (b)-(d) to valley $K'$.
\label{fig:Dyn1}}
\end{figure} 
We considered two limiting cases for the incidence angle: (1) horizontal incidence where the wavepacket is originally centered at $(x_0,y_0)=(-150,0)\text{ nm}$ and moves with momentum and pseudospin polarization oriented along the $+x$-axis (for valley $K$, for valley $K'$ the pseudospin is reversed accordengly with eq.(\ref{eq:statesK}) ); and (2)
vertical incidence where the wavepacket is originally centered at $(x_0,y_0)=(0,-150)\text{ nm}$ and moves with momentum and pseudospin polarization oriented along the $+y$-axis. In our implementation,  we used a time step of $\Delta t=0.05\text{ fs}$ and let it evolve in time from $t_i=0$ up to $t_f=700\text{ fs}$, taking a spatial squared region of $L=1500\text{ nm}$ with a two dimensional mesh given by $\Delta L=0.1\text{ nm}$. 

For illustration, in Fig.~\ref{fig:Dyn1} we choose horizontal incidence and present plots of the probability density ($|\Psi(x,y)|^2$)for each valley, with and without considering the scalar field . All the plots show the spatial distribution of the probability density per valley at time $t=350\text{ fs}$, once the wavepacket has left the region where the intensity of the pseudomagnetic field concentrates ($r<100\text{ nm}$). Notice that horizontal incidence corresponds to $\theta_1=180^\circ$ in Fig.~\ref{fig:sd}. 

\begin{figure}[!htbp]
\begin{center}
\includegraphics[scale=0.20]{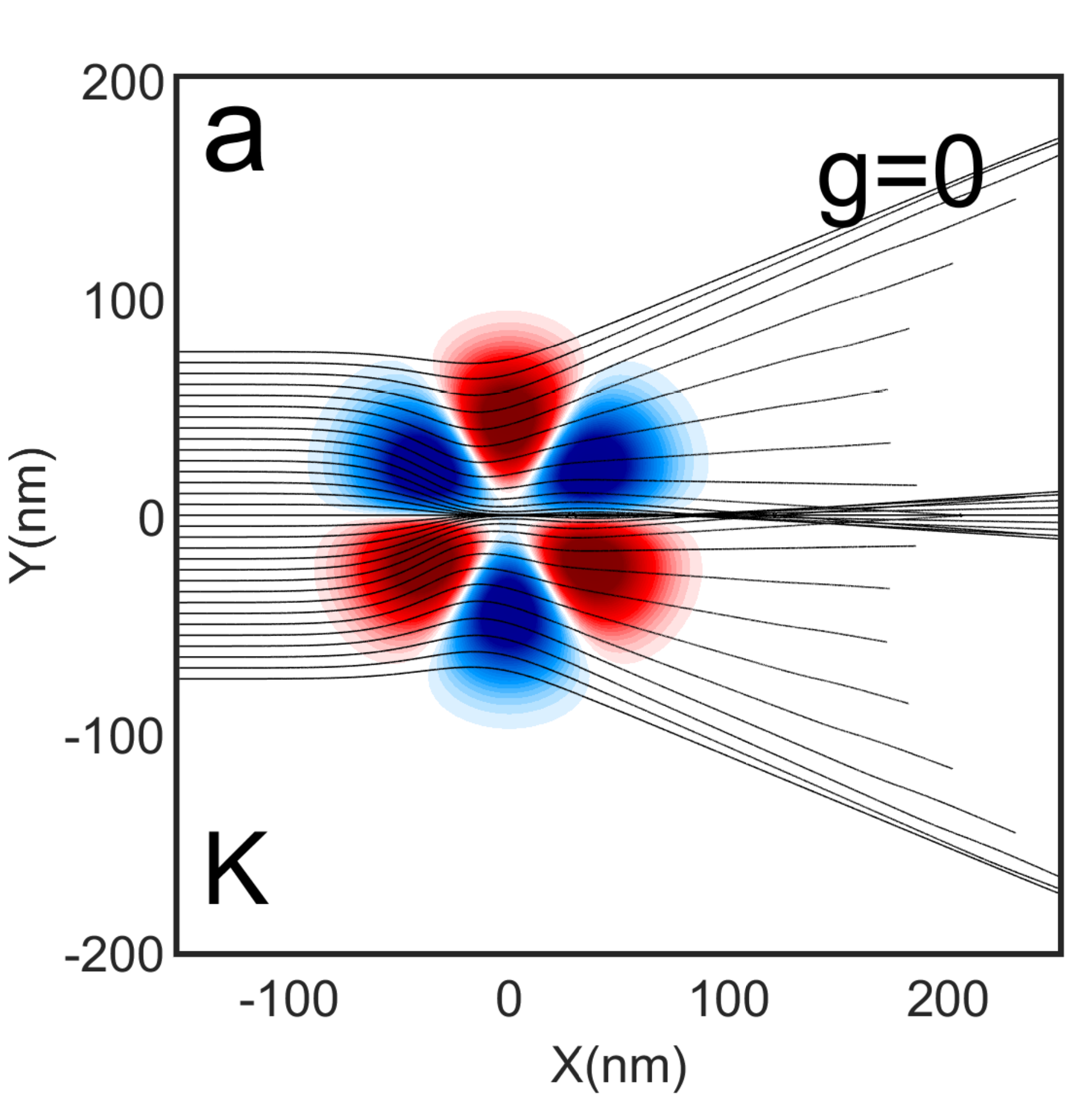}
\includegraphics[scale=0.20]{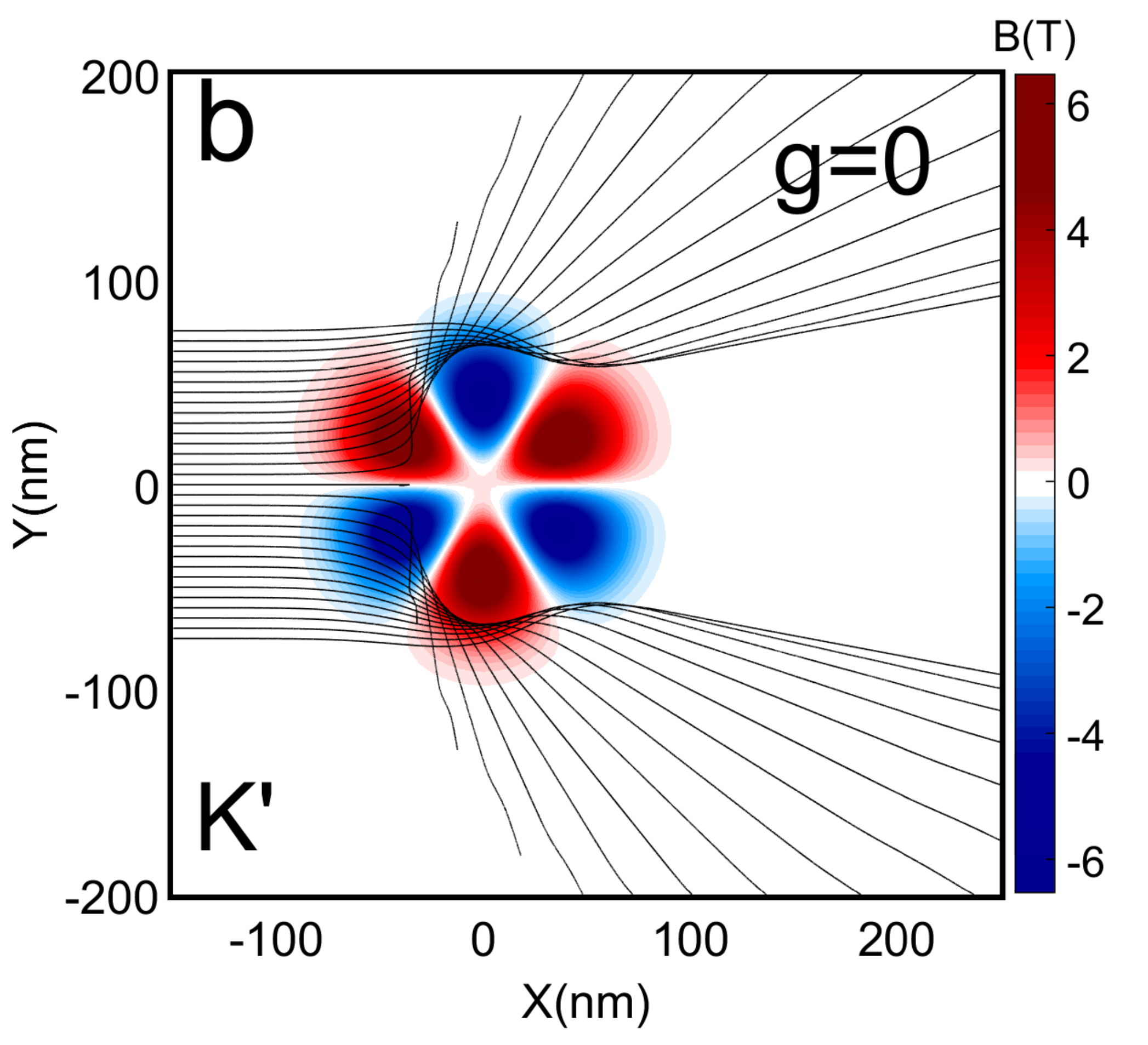}
\includegraphics[scale=0.20]{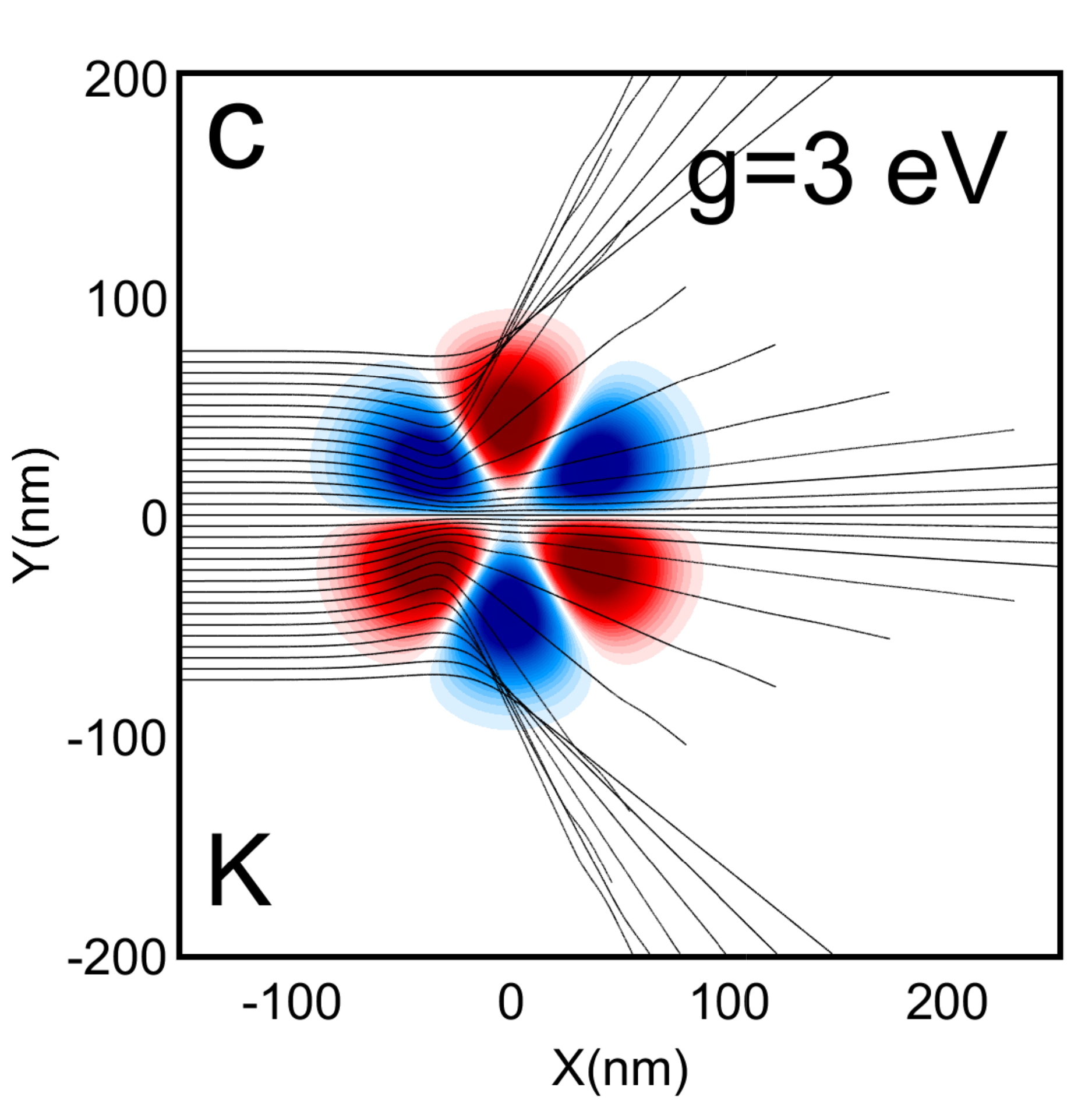}
\includegraphics[scale=0.20]{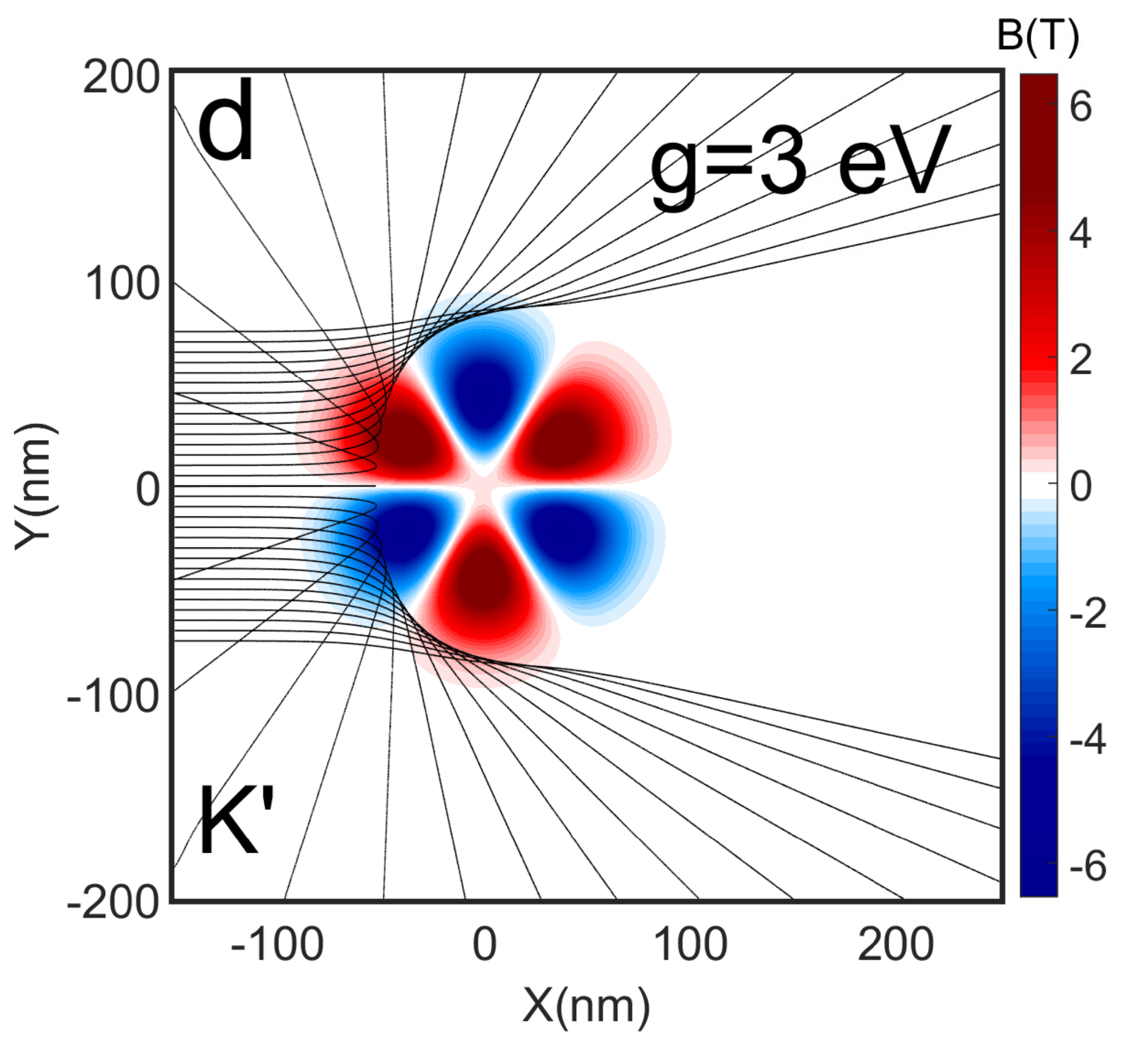}
\end{center}
\caption {(Color online) Trajectories of $\left\langle\boldsymbol{r}\right\rangle$ (black) of an incident wave packet coming from the left with $E=110\text{ meV}$ for different impact parameters plotted on top of the pseudomagnetic field profile for valleys $K$ (left column) and $K'$ (right column) with $g=0$ (a-b) and $g=3\text{ eV}$ (c-d).
\label{fig:horizontal}}
\end{figure}
Fig.~\ref{fig:Dyn1}(a) and Fig.~\ref{fig:Dyn1}(b) shows the probability density with vanishing scalar field ($g=0$). We observe that the scattering effects are dramatically different for each valley, producing regions of concentration of  the probability density that are valley asymmetric.  The physics behind this behavior lies in the geometrical distribution of the field (see Fig. \ref{fig:fields}) and the fact that the incident valley-$K$ polarized electron wave packet experiences an effective non-uniform magnetic field that takes the opposite sign for the incident valley-$K'$ \cite{settnes2016valley}. For the former case, the electron wave packets are essentially guided and focused through snake states\cite{Peeters2016valley} that surround regions with opposite pseudomagnetic fields (see Fig.~\ref{fig:fields}a)  and finally transmitted away to the right from the deformation region. However, in the later case (valley $K'$) the opposite sign for the pseudomagnetic field pushes away the wave packet from the center of the deformation. This reflects a strong  perpendicular scattering at angles between $90^\circ$
and $270^\circ$ with respect to the incidence direction, [see Fig.~\ref{fig:Dyn1}b], whiles scattering is almost absent for valley $K$ [see Fig.~\ref{fig:Dyn1}a]. 

When the scalar field is present, Fig.~\ref{fig:Dyn1}(c) and Fig.~\ref{fig:Dyn1}(d), the valley asymmetric scattering persist but the wave-packet profiles for each valley changes. Particularly, for the valley $K'$ we can see a strong backscattering making the Gaussian deformation basically transparent for valley $K$ and reflective for $K'$ for normal incidence. The presence of backscattering is in qualitative agreement with the results of the Born approximation.  See for instance the valley polarization efficiency plot in Fig.~\ref{fig:pol} and the results of the wavepacket dynamics shown in Fig.~\ref{fig:Dyn1}c and Fig.~\ref{fig:Dyn1}d.

Classical studies of scattering usually include the calculation deflection angle as a function of the impact parameter. In our case, we define the deflection angle as the angle between the incoming and outgoing direction, using the trajectory on the expected value of the position operator\cite{schliemann2008cyclotron,rakhimov2011} $\left\langle\boldsymbol{r}\right\rangle$.  Classical trajectories of the wave packet for horizontal incidence (from  the left) to the pseudomagnetic field region produced by the bump for both valleys,  are shown in Fig.~\ref{fig:horizontal} (panels a,b without the presence of the scalar field, and c, d with it).
\begin{figure}[!htbp]
\begin{center}
\includegraphics[scale=0.20]{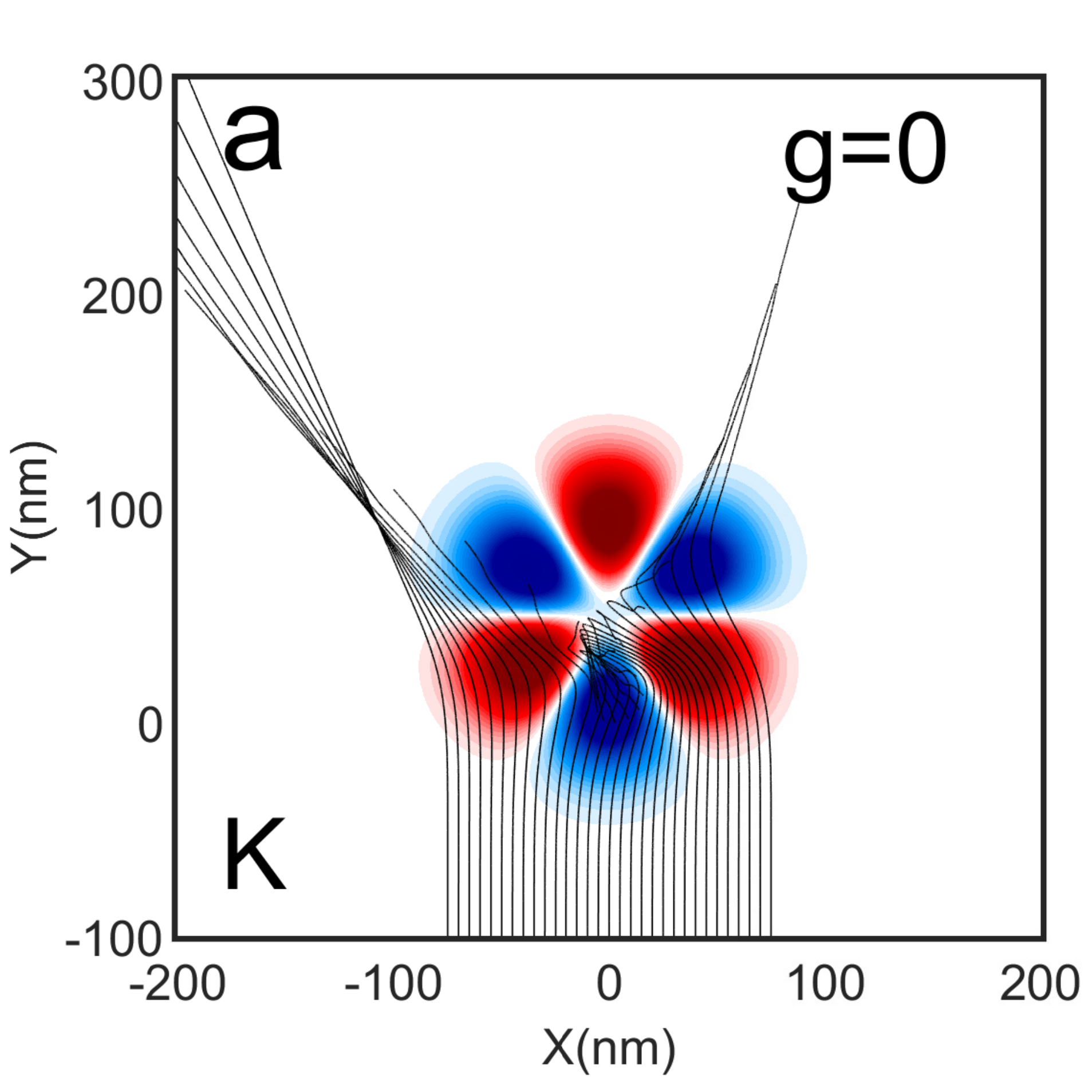}
\includegraphics[scale=0.20]{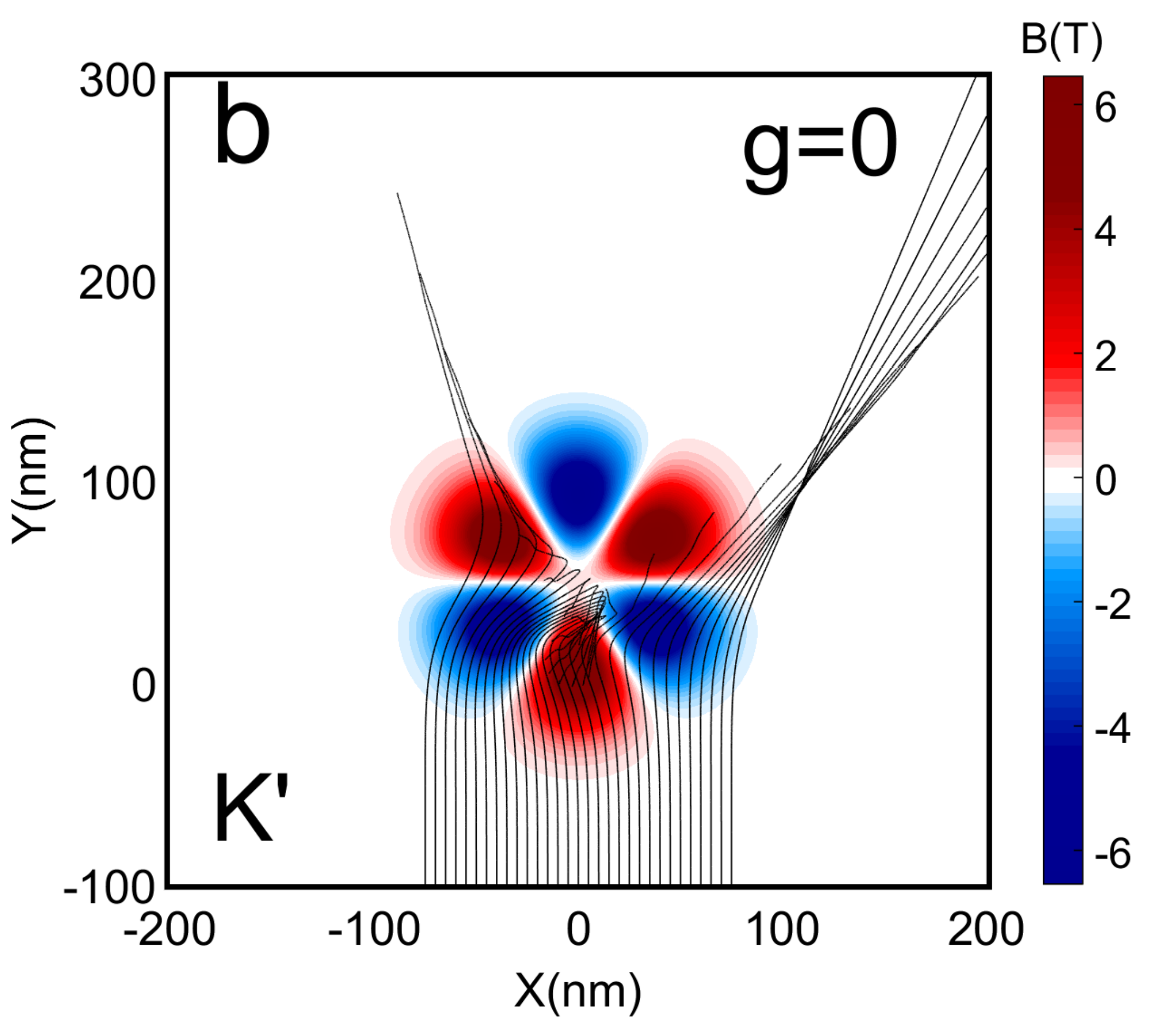}
\includegraphics[scale=0.20]{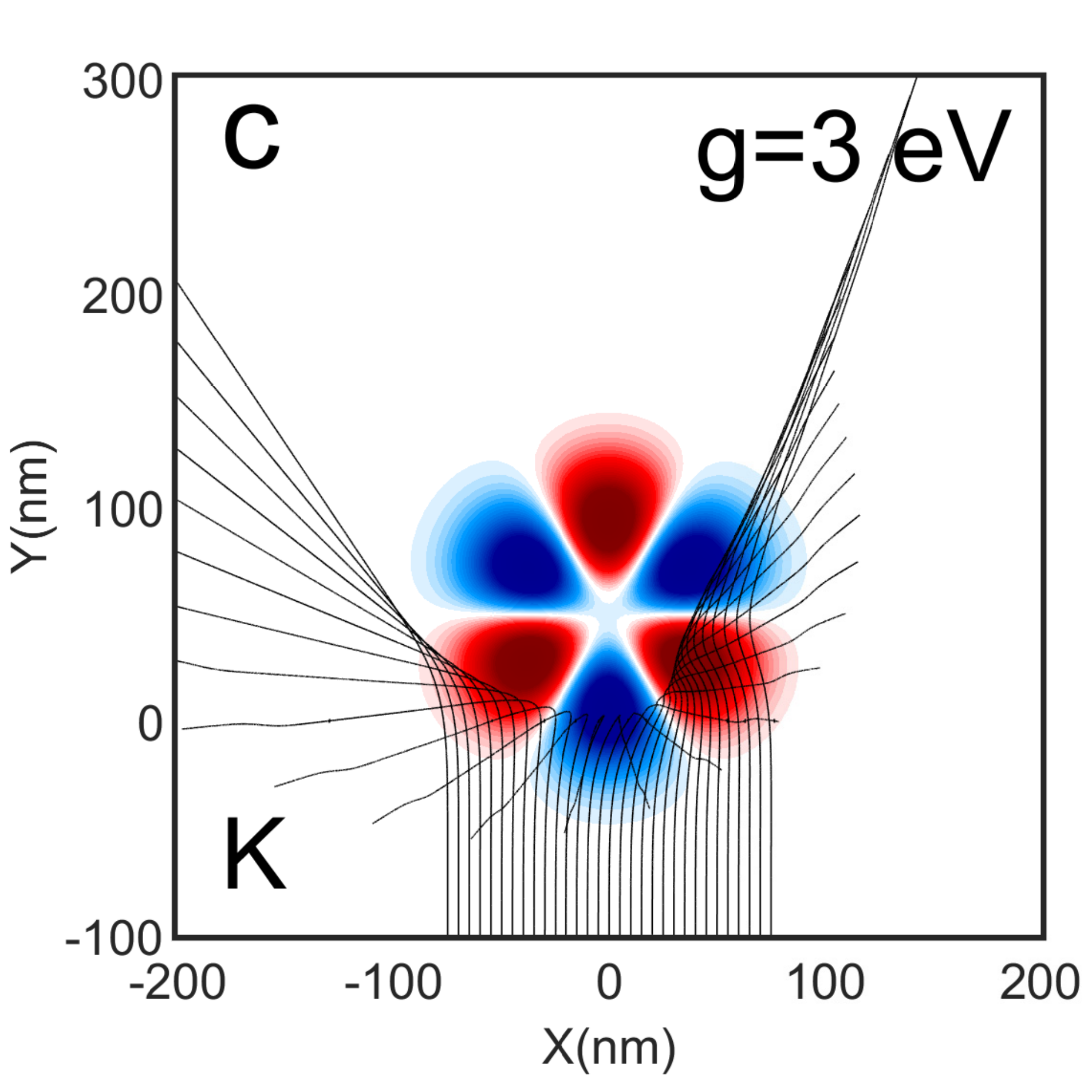}
\includegraphics[scale=0.20]{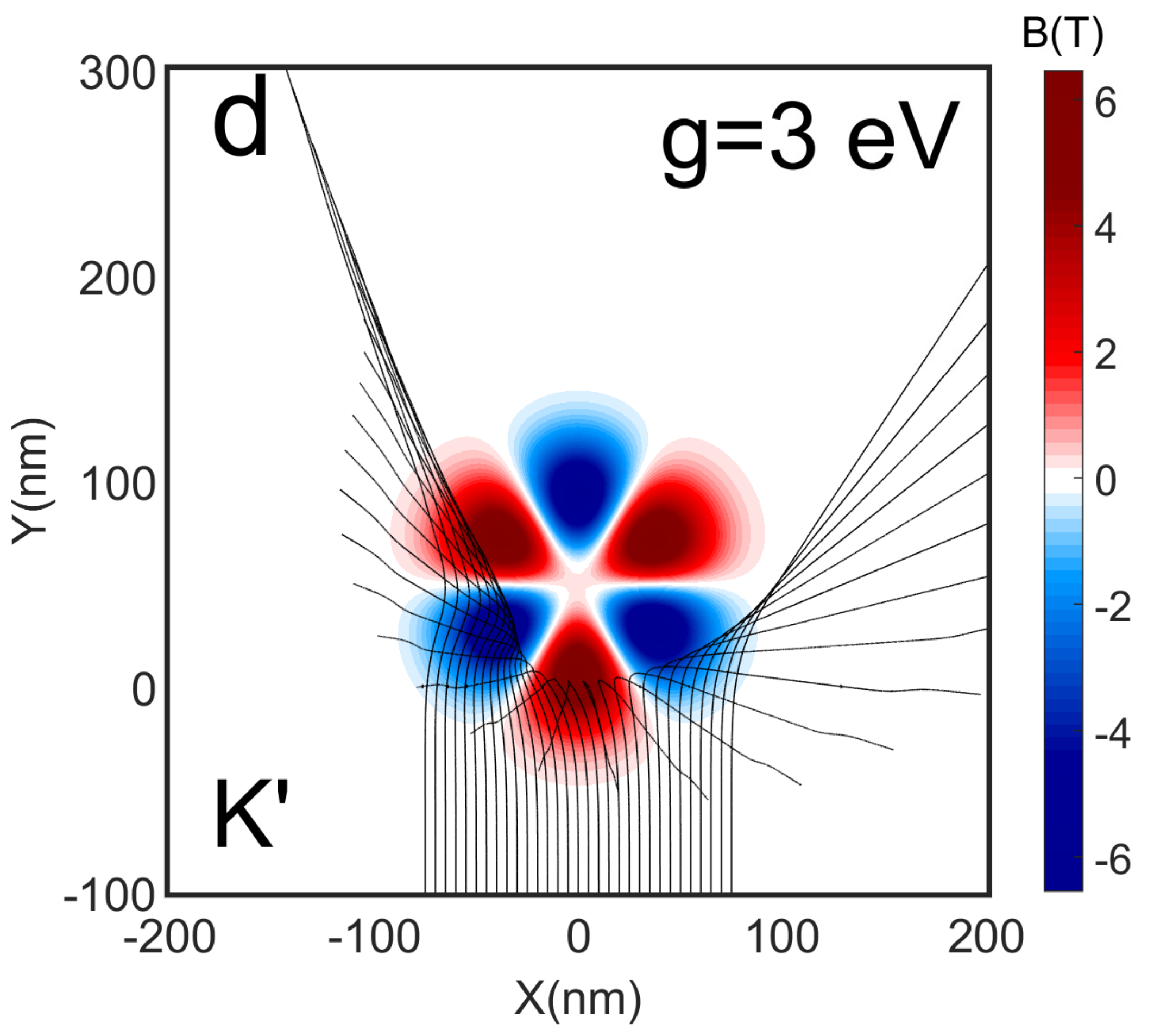}
\end{center}
\caption {(Color online)Trajectories of $\left\langle\boldsymbol{r}\right\rangle$ (black) of an incident wave packet coming from the bottom with $E=110\text{ meV}$ for different impact parameters plotted on top of the pseudomagnetic field profile for valleys $K$ (left column) and $K'$ (right column) with $g=0$ (a-b) and $g=3\text{ eV}$ (c-d).
\label{fig:vertical}}
\end{figure}
Explicitly we take a Gaussian wave packet initially centered at $(-150,y_0)\text{nm}$ moving (with average wave number $\boldsymbol{k_o}=k_{ox} \hat x$) towards the locally strained region. For this setup, $y_0$ defines the impact parameter. Black curves correspond to the average trajectories with different values of the impact parameter $y_0=\left\lbrace-75,70,65...,75\right\rbrace$ nm.  We observe an opposite behavior of the classical trajectories when comparing the cases for the $K$ and $K'$ valleys. While for valley $K'$ the bump acts -in terms of geometrical optics arguments- as a divergent pseudomagnetic lens, for valley $K$ it behaves as a convergent lens.
For instance, the case for $g=0$ shows a focusing of the stream of electrons to a narrow region (Fig.~\ref{fig:horizontal}~a) for valley $K$, whereas it shows deflecting trajectories in a bifurcated pattern at $\pm y$ direction for valley $K'$ (Fig.~\ref{fig:horizontal}~b). Notice that in the first scenario the classical trajectories penetrates the whole distorted region whiles in the second case experiences a deflection, avoiding the bump region. Therefore the locally strain region  will yield preferential directions of valley polarization, as being discussed in the literature\cite{settnes2016valley,Peeters2016valley,stegmann2016valley}. 
The overall behavior of the classical trajectories remains unchanged in the presence of the scalar field, as shown in Fig.~\ref{fig:horizontal}~d-f. However, in this case the deflection angles for valley $K$ are greater, with well defined directions of valley polarization as well. 
Other incident directions may also offer valley splitting properties, in particular when the incidence is directed towards one of the lobules of pseudomagnetic field, (see Fig.~\ref{fig:vertical}). 

\begin{figure}[!htbp]
\begin{center}
\includegraphics[scale=0.20]{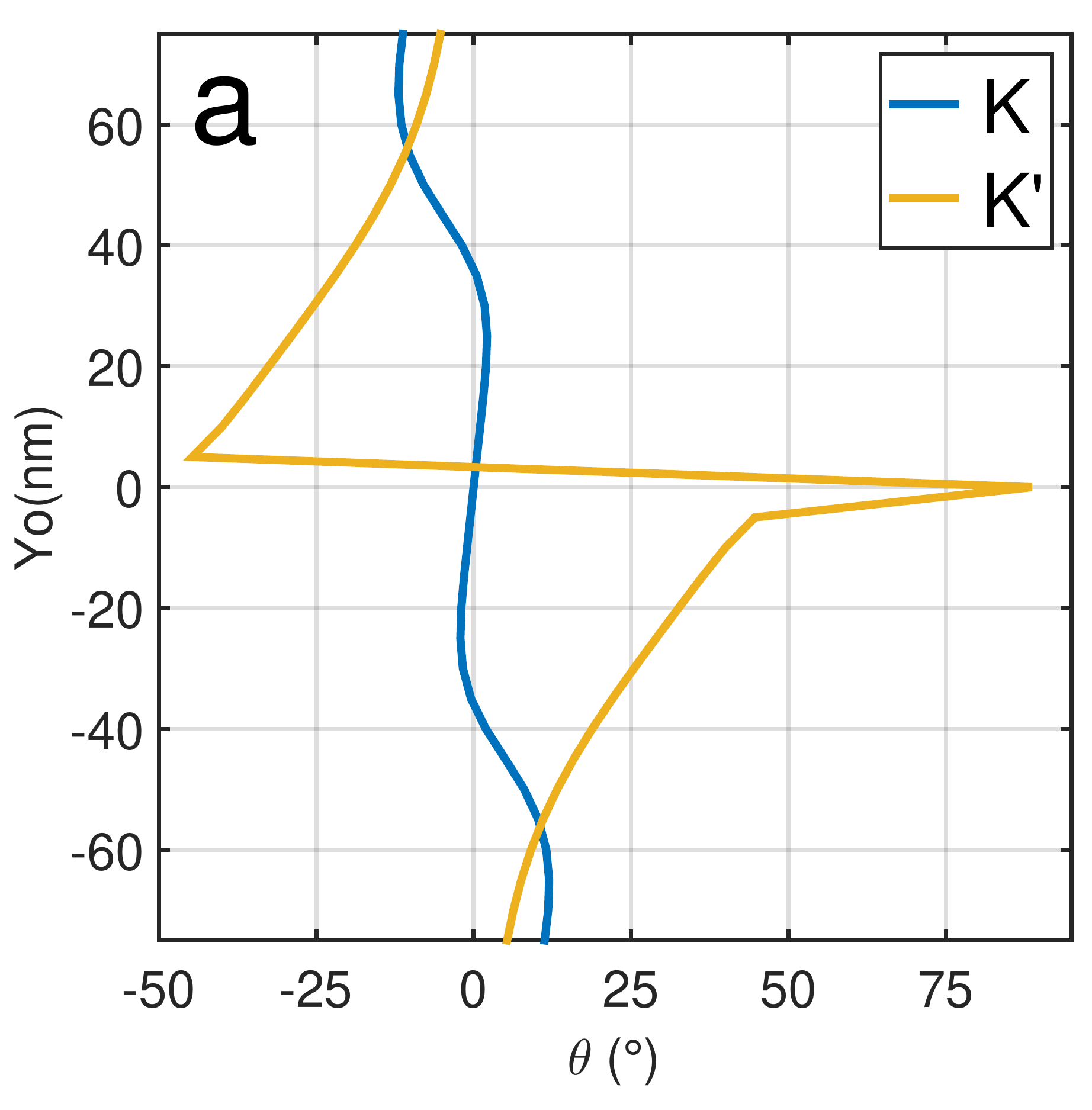}
\includegraphics[scale=0.20]{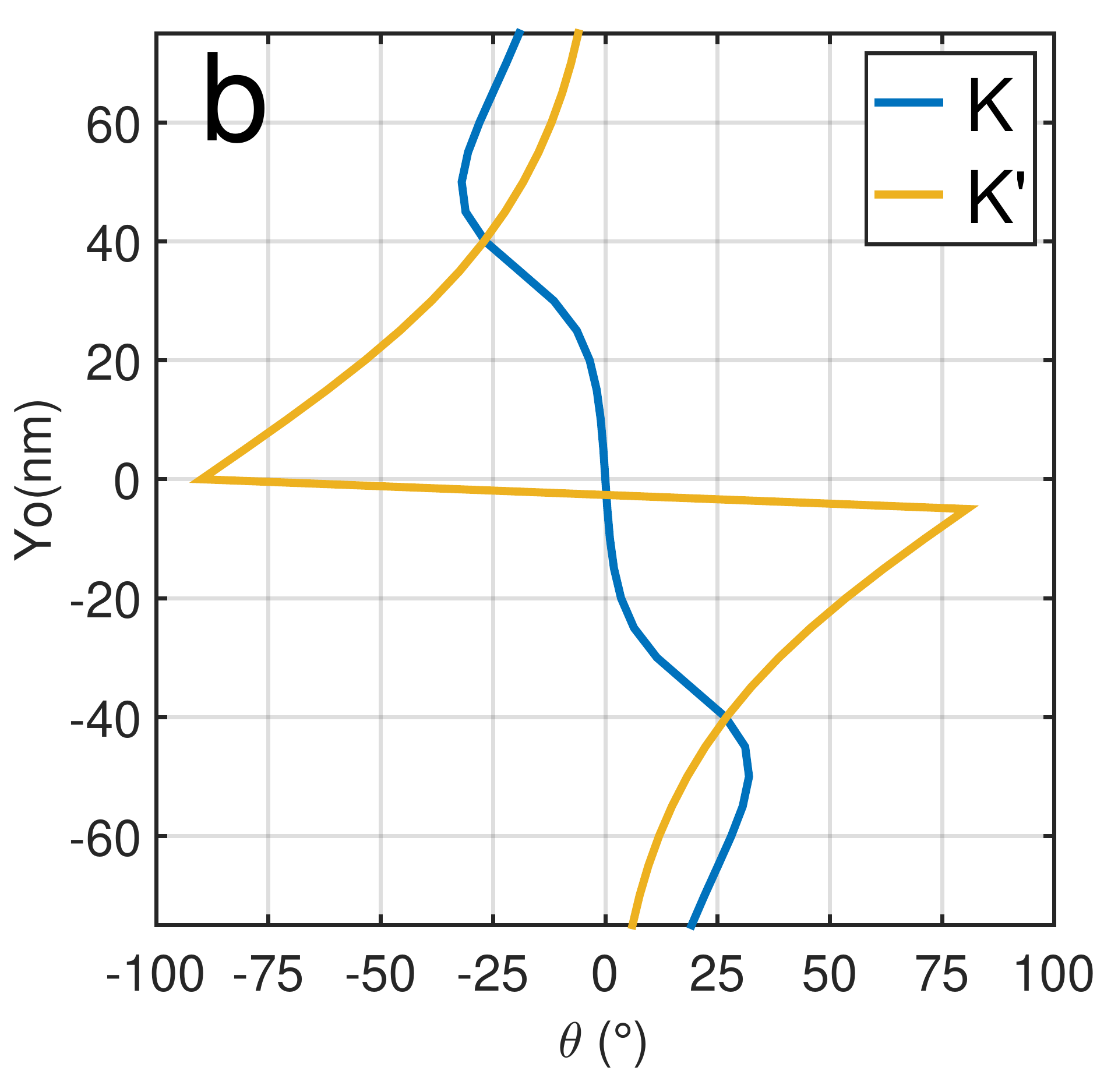}
\includegraphics[scale=0.20]{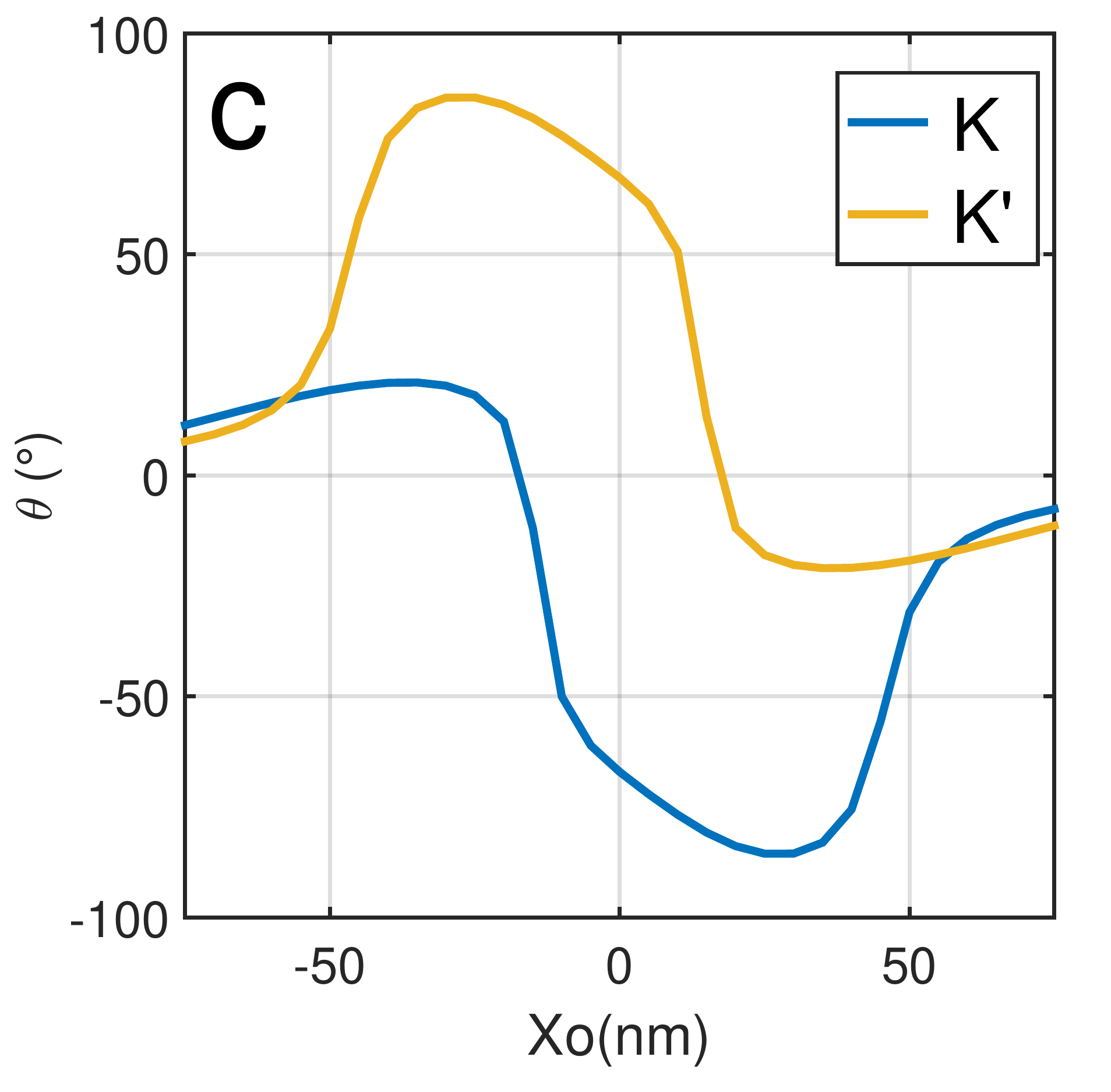}
\includegraphics[scale=0.20]{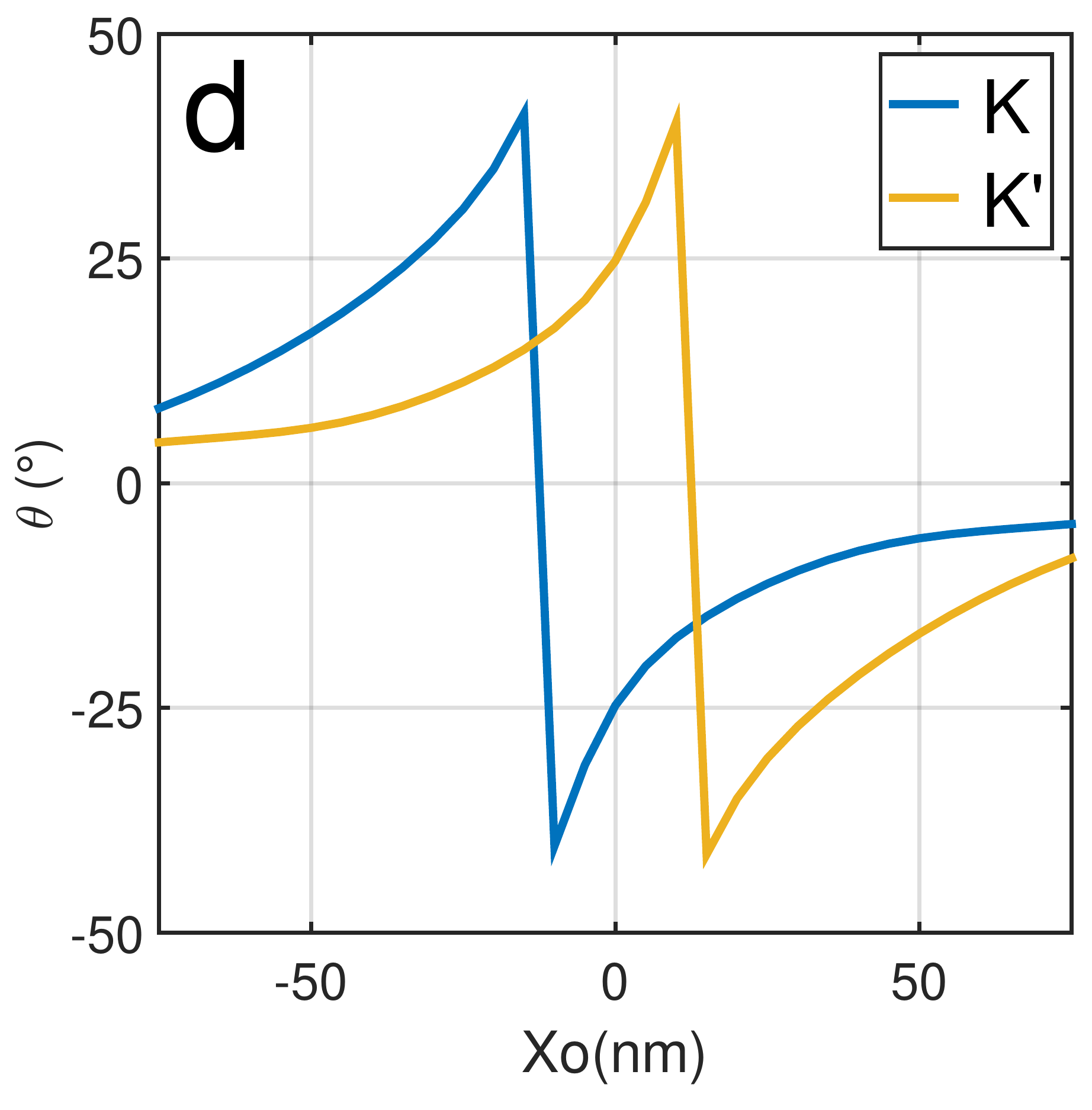}
\end{center}
\caption {(Color online) Relation between the deflection angle and the impact parameter $Y_0$ ($X_0$) for horizontal (vertical) incidence for valley $K$ [blue curve] and valley $K'$ [yellow curve]. Cases without scalar field are shown in panel a (c),  and cases with finite scalar field are shown in panel b (d).
\label{fig:deflection}}
\end{figure}
This is also consistent  with the results within the Born approximation when the scalar field is present (see Fig.\ref{fig:sd}d); vertical incidence produce directions around $60^\circ$ and $120^\circ$ degrees with high valley polarization.

Finally we explore the relation between the deflection angle $\theta$ and the impact parameter for horizontal (Fig.~\ref{fig:horizontal}) and vertical (Fig.~\ref{fig:vertical})  incidences and are depicted in Fig.~\ref{fig:deflection}. In the case of horizontal incidence, we call the attention to the fact that there is a small range of values for the impact parameter  ($Y_0\approx b/10$) where one valley component ($K$) is almost not deflected, while the other valley ($K'$) presents two maximal values of deflection around zero, the latter occurs in both situations with (Fig.~\ref{fig:deflection}~a) and without (Fig.~\ref{fig:deflection}~b) scalar field.  This valley asymmetric behavior of the deflection angle is more pronounced in the case of vertical incidence (Fig.~\ref{fig:deflection}~c-d), where for small impact parameters ($x_0\approx b/10$) each valley component is directed towards opposites directions. More interesting is the fact that for bigger impact parameters ($b/10<x_0<b/10$) the valley changes when the scalar field is present, making possible to control this degree of freedom.

\section{Conclusions}

Using the low energy approximation to describe the interaction between deformations  and electrons moving in a graphene membrane, we have described the role of the scalar field in the ability of Gaussian bumps to generate valley polarization and valley splitting/polarization in graphene systems. Our results were obtained using an analytical approach based on the Born approximation. In addition, we  characterize the valley asymmetric scattering by introducing a valley polarization efficiency, $\cal{P}$, that clearly shows the polarization effects. Similar effects are predicted for centrosymmetric external scalar fields. We also use  a dynamical approach and studied the wave-packet dynamics of an encounter with the pseudomagnetic profile caused by a Gaussian bump. 
We present  results for the average trajectories of wave packets in locally strained graphene that clearly shows the enhancement of the wave packet focusing and beam splitting effects when the scalar field is present.We have shown that a simple bump geometry in graphene and considering effects of the scalar field can promote  valley current flowing with opposite directions. Quite remarkable, we find that there is also the possibility of tuning the valley splitting effects
solely by electrical means in combination with strain fields. We believe that these results can be exploited in the implementation of valleytronic devices.

\section*{Acknowledgements}
R.C. acknowledges usefull discussions with D. Faria, M. Asmar, N. Sandler, and G Naumis, as well as constructive criticism by M. Vozmediano and S. Barraza Lopez. R. C. acknowledges the support of PRODEP.  F.M., S.Z. and R.C. acknowledges the support of PAPIIT-UNAM through the project IN111317. S.Z. was partially supported by PROMEP-DGEST Beca No. 022007014.

\appendix
\section{Evaluation of integrals in Eq.\ref{eq:Fn}}\label{sec:F2y0}
We departure from the integral formula\cite{abramowitz1964handbook},
\begin{equation}\label{eq:Fn-abramowitz}
\int_0^{\infty}e^{-a^2t^2}t^{\nu+1}J_{\nu}(bt)dt=\dfrac{b^\nu}{(2a^2)^{\nu+1}}e^{-\dfrac{b^2}{4a^2}}
\end{equation}     
where $J_\nu$ is the Bessel function of order $\nu$. With this formula we can evaluate Eq.(\ref{eq:Fn}) for $n=2$. In order to evaluate the integral when $n=0$ we can derivate both sides of Eq.(\ref{eq:Fn-abramowitz}) with respect of $a$, to obtain
\begin{equation}\label{eq:Fm-abramowitz}
\int_0^{\infty}e^{-a^2t^2}t^{\nu+3}J_{\nu}(bt)dt=\dfrac{2b^\nu}{(2a^2)^{\nu+2}}\left[\nu+1-\dfrac{b^2}{4a^2}\right]e^{-\dfrac{b^2}{4a^2}}
\end{equation}   

\section{Numerical Methodology}
 We start by writing the Dirac Hamiltonian for each valley in the following way,  
\begin{equation}\label{eq:Hk}
H_{\eta}=v_{F}\boldsymbol{\sigma}_{\eta}\cdot \left(\hat{\boldsymbol{p}}-\eta\bm{\mathcal A}(\boldsymbol{r})\right) +V(\boldsymbol{r})=H_{\eta}^{o}+U_{\eta}(\boldsymbol{r}),
\end{equation}
\noindent where the term 
\begin{equation}\label{eq:Ho}
H^{o}_\eta=v_{F}\boldsymbol{\sigma}_{\eta}\cdot \hat{\boldsymbol{p}}
\end{equation}
is the bare Hamiltonian for graphene (without strains), at the valley $K$  ($\eta=1$) or $K'$ ($\eta=-1$), and depends only on the momentum operator, whereas the strain and scalar potential part is carried by $U_{\eta}(\boldsymbol{r})$, given by Eq.\,(\ref{eq:H1}). Note that $\left[H^{o}_\eta, U_{\eta}(\boldsymbol{r})\right]\neq 0$, nevertheles, the corresponding time evolution operator ${\cal U}_{\eta}(t)=exp(-iH_{\eta} (t-t_o)/\hbar)$ can be approximated using the standard time-splitting spectral method that consists in a second order  Trotter decomposition of the evolution operator at any given time step $\Delta t$~\cite{suzuki1990fractal,Peeters2010waveTB}
\begin{equation}\label{eq:TEO}
{\cal U}_{\eta}(t)\approx e^{-iU_{\eta}\Delta t/2\hbar}e^{-iH^{o}_\eta\Delta t/\hbar}e^{-iU_{\eta}\Delta t/2\hbar} +{\cal O}(\Delta t^3),
\end{equation}  
\noindent which conveniently decomposes the application of the time-evolution operator in kinetic and potential terms. Then the wave function $\psi_{\eta}(t+\Delta t)$ can be obtained in terms of $\psi_{\eta}(t)$ by the  application of the time evolution operator as follows,
\begin{equation}\label{Psi}
\Psi_{\eta}(t+\Delta t)\simeq e^{-iU_{\eta}\Delta t/2\hbar}e^{-iH^{0}_\eta\Delta t/\hbar}e^{-iU_{\eta}\Delta t/2\hbar}\Psi_{\eta}(t),
\end{equation}

\noindent which is correct  up to second order in $\Delta t$. 
Note that the terms within $H^o_\eta$  do not commute with each other, and neither the terms within $U_\eta$ as $\left[\sigma_x,\sigma_y \right]=2i\sigma_z$. Thus to avoid diagonalization at each time step, it is convenient to split Eq.\,(\ref{Psi}) even further. In order to do this we employ the  Zassenhaus formula instead, which establish that for any two linear noncommutative $X$ and $Y$ operators in the Lie algebra
%
\begin{equation}\label{eq:Zassenhaus}
e^{t\left(X+Y\right)}=e^{tX}e^{tY}e^{-\frac{t^{2}}{2}\left[X,Y\right]}e^{\frac{t^{3}}{6}(2\left[Y,\left[X,Y\right]\right]+\left[X,\left[X,Y\right]\right])}...
\end{equation}
\noindent in which the exponents of higher order in $t$ are likewise homogeneous Lie polynomials (nested commutators). Thus we can approximate the time evolution operator as a sequential product of exponential terms of the form $e^{i\hat{A}\sigma_{\mu}}$
\noindent where $\hat{A}$ is an operator that depends either on momentum or the position and $\sigma_{\mu}=\{\sigma_0,\sigma_x,\sigma_y,\sigma_z,\}$. When $\hat{A}$ is position dependent only its application is straightforward, but when it depends on momentum we use  the Cayley's expansion 
\begin{equation}
e^{i\hat{A}\sigma_{\mu}} \simeq\left( \dfrac{1+\frac{i}{2}\hat{A}\sigma_{\mu} } {1-\frac{i}{2}\hat{A}\sigma_{\mu}}\right) +{\cal{O}}(\hat{A}^2)
\end{equation}
for the exponentials to ensure unitarity and particle conservation at each time step.  

\bibliographystyle{apsrev4-1}
\bibliography{biblio.bib}
\end{document}